\documentclass[twocolumn,superscriptaddress,secnumarabic,
amssymb,amsmath,nobibnotes,aps,prd,showpacs,nofootinbib]{revtex4}%
\usepackage{graphicx}
\usepackage{epsf}
\usepackage{bm}
\usepackage{amsmath}
\usepackage{amsfonts}
\usepackage{amssymb}
\usepackage{epstopdf}
\usepackage{natbib}%
\usepackage{rotating}
\usepackage{pdflscape}
\usepackage{adjustbox}
\providecommand{\U}[1]{\protect\rule{.1in}{.1in}}
\newcommand{\be}{\begin{equation}}
\newcommand{\ee}{\end{equation}}

\newcommand{\mincir}{\raise
-3.truept\hbox{\rlap{\hbox{$\sim$}}\raise4.truept\hbox{$<$}\ }}
\newcommand{\magcir}{\raise
-3.truept\hbox{\rlap{\hbox{$\sim$}}\raise4.truept\hbox{$>$}\ }}

\usepackage{color}

\begin{document}

\title{\textit{All-inclusive} interacting dark sector cosmologies}

\author{Weiqiang Yang}
\email{d11102004@163.com}
\affiliation{Department of Physics, Liaoning Normal University, Dalian, 116029, People's Republic of China}

\author{Eleonora Di Valentino}
\email{eleonora.divalentino@manchester.ac.uk}
\affiliation{Jodrell Bank Center for Astrophysics, School of Physics and Astronomy, 
University of Manchester, Oxford Road, Manchester, M13 9PL, UK.}

\author{Olga Mena}
\email{omena@ific.uv.es }
\affiliation{IFIC, Universidad de Valencia-CSIC, 46071, Valencia, Spain}

\author{Supriya Pan}
\email{supriya.maths@presiuniv.ac.in}
\affiliation{Department of Mathematics, Presidency University, 86/1 College Street, Kolkata 700073, India}

\author{Rafael C. Nunes}
\email{rafadcnunes@gmail.com}
\affiliation{Divis\~{a}o de Astrof\'{i}sica, Instituto Nacional de Pesquisas Espaciais, Avenida dos Astronautas 1758, S\~{a}o Jos\'{e} dos Campos, 12227-010, SP, Brazil}

%%%%%%%%%%%%%%%%%%%%%%%%%%%%%%%%%%%%%%%%%%%%%%%%%%%%%%%%%%%%%%%%%%%%%%%%%%%%%%%%%%%%%%%%%%%%%%%%%%%%%%%%%%%%%%%%%%%%%%%%%%%%%%%%

\begin{abstract}
In this paper we explore possible extensions of Interacting Dark Energy cosmologies, where Dark Energy and Dark Matter interact non-gravitationally with one another. In particular, we focus on the neutrino sector, analyzing the effect of both neutrino masses and the effective number of neutrino species. We consider the Planck 2018 legacy release data combined with several other cosmological probes, finding no evidence for new physics in the dark radiation sector. The current neutrino constraints from cosmology should be therefore regarded as robust, as they are not strongly dependent on the dark sector physics, once all the available observations are combined. Namely, we find a total neutrino mass $M_\nu<0.15$~eV and a number of effective relativistic degrees of freedom of $N_{\rm eff}=3.03^{+0.33}_{-0.33}$, both at 95\%~CL, which are close to those obtained within the $\Lambda$CDM cosmology, $M_\nu<0.12$~eV and $N_{\rm eff}=3.00^{+0.36}_{-0.35}$ for the same data combination.
\end{abstract}

%%%%%%%%%%%%%%%%%%%%%%%%%%%%%%%%%%%%%%%%%%%%%%%%%%%%%%%%%%%%%%%%%%%%%%%%%%%%%%%%%%%%%%%%%%%%%%%%%%%%%%%%%%%%%%%%%%%%%%%%%%%%%%%%
\maketitle
%%%%%%%%%%%%%%%%%%%%%%%%%%%%%%%%%%%%%%%%%%%%%%%%%%%%%%%%%%%%%%%%%%%%%%%%%%%%%%%%%%%%%%%%%%%%%%%%%%%%%%%%%%%%%%%%%%%%%%%%%%%%%%
%\myclassification{98.80.Cq, 98.80.-k}
%%%%%%%%%%%%%%%%%%%%%%%%%%%%%%%%%%%%%%%%%%%%%%%%%%%%%%%%%%%%%%%%%%%%%%%%%%%%%%%%%%%%%%%%%%%%%%%%%%%%%%%%%%%%%%%%%%%%%%%%%%%%%
\section{Introduction}
\label{sec:intro}

There are overwhelming observations which confirm that our universe is undergoing through a phase of accelerated expansion. Within the context of general relativity, this effect can be described by adding a dark energy component (DE), characterized by a negative pressure. According to the observational evidences, nearly 68\% of the total energy budget of the universe is filled up with such a dark energy fluid. Observations further predict that around 28\% of the total energy budget of the universe corresponds to non-luminous dark matter (DM)~\cite{Aghanim:2019ame,Aghanim:2018eyx,Aghanim:2018oex}. Observations from different astronomical sources seem to point to a DE component very similar to a cosmological constant, and to a pressure-less DM component, also known as cold dark matter (CDM). The former two components are the basic ingredients of the so-called $\Lambda$CDM cosmology. However, given the fact that \textit{(a)} the precise nature of the DE and the DM fluids remains unknown, despite the large number of devoted measurements to unravel the underlying physics~\cite{Suzuki:2011hu,Crocce:2015xpb,Alam:2015mbd,Hinshaw:2012aka,Ade:2015xua}; and, \textit{(b)} due to a number of persisting tensions within the minimal $\Lambda$CDM scheme, there is still plenty of room for other possible scenarios with non-minimal dark sector physics. In this regard, we shall consider here interacting dark sector cosmologies, in which the DE and the DM components interact non-gravitationally~\cite{Billyard:2000bh,Barrow:2006hia,Amendola:2006dg,He:2008tn,Valiviita:2008iv,Gavela:2009cy,Majerotto:2009np,Gavela:2010tm,Clemson:2011an,Pan:2013rha,Yang:2014vza,Yang:2014gza,Pan:2012ki,Pan:2016ngu,Mukherjee:2016shl,Sharov:2017iue,Yang:2017yme,Yang:2017zjs,Yang:2017ccc,Pan:2017ent,Yang:2018pej,Yang:2018ubt,Yang:2018xlt,Martinelli:2019dau,Paliathanasis:2019hbi,Pan:2019jqh,Yang:2019bpr,Yang:2019vni,Papagiannopoulos:2019kar,Kumar:2017dnp,DiValentino:2017iww,Yang:2018euj,Yang:2018uae, Kumar:2016zpg,Kumar:2017bpv,Kumar:2019wfs,Yang:2019uog}. Indeed, the dynamics of such a universe could have been present since very early times, modifying other dark sector physics, such as that of dark radiation, i.e. that of the neutrinos. Therefore, the question we would like to address here is the following: \textit{are cosmological limits on dark radiation standard physics strongly dependent on the dark sector modelling ?}

In order to investigate this above issue we have used a set of observational data from various important cosmological sources. This set includes the cosmic microwave background radiation, baryon acoustic oscillations distance measurements, local measurement of the Hubble constant from Hubble Space Telescope, Pantheon sample of Supernovae Type Ia and finally the Hubble parameter measurements at different redshifts from the cosmic chronometers. 

The structure of the paper is as follows. Section~\ref{sec-2} introduces the cosmological model explored here while  Sec.~\ref{sec-3} describes the methodology and the measurements exploited in our data analyses. Section~\ref{sec-4} presents our results and we conclude the article in Sec.~\ref{sec-5}.

\section{Interacting dark universe}
\label{sec-2}

We consider an interacting dark sector scenario between cold DM and DE in a homogeneous and isotropic flat universe where the gravitational sector is described by the Einstein's general relativity. The energy-momentum tensor conservation for cold DM and DE leads to the coupled equations:
\begin{eqnarray}
\label{cons-dark2}
&&\dot{\rho}_c + 3 H \rho_c =-Q,\\
&&\dot{\rho}_{x} + 3 H (1+w_{x}) \rho_{x} = Q~,
\end{eqnarray}
where we have explicitly used the DE state parameter $w_{x}  = p_{x}/\rho_{x}$ (here $p_{x}$ is the pressure of the DE fluid and  $\rho_x$ is its energy density) and similarly for DM we use its pressure, $p_c$, which, for a CDM component is $0$ and $\rho_c$ as its energy density. Note that here $H$ denotes the Hubble rate of the prescribed homogeneous and isotropic universe, that means the Friedmann-Lema\^{i}tre-Robertson-Walker (FLRW) universe. 
In the above equation (\ref{cons-dark2}) we have introduced the quantity $Q$, which is known as the interaction rate between the dark sectors. While the exact form of the interaction rate is not known, possible phenomenological descriptions at classical and quantum levels have been developed in the literature (see Ref.~\cite{Gleyzes:2015pma} for a comprehensive review), trying to recover some different interaction models from scalar field theories, see e.g. the recent article \cite{Pan:2020zza} or from other cosmological contexts, see e.g. \cite{Pan:2020mst}. 
In this present work, we restrict ourselves to most well known parametric form for the coupling function $Q$: 
\begin{eqnarray}
&& Q = 3 H \xi \rho_{x},\label{model}
\end{eqnarray}
where $\xi$ is the coupling parameter that characterizes the strength of the coupling. Where, in the convenience of this paper, the condition $\xi < 0$ corresponds to the energy flow from DE to DM, and  $\xi > 0$ represents the opposite case.  We refer to Refs.~\cite{Yang:2017ccc, Yang:2018euj, Yang:2018xlt, Yang:2018uae,Pan:2019jqh} 
for details concerning the linear perturbation theory within these coupled cosmologies. 

\section{Observational data and methodology}
\label{sec-data+analyses}
\label{sec-3}
We briefly describe below the observational data sets that we have used in the present work. 
\begin{itemize}
    \item {\bf Cosmic Microwave Background(CMB)}: we make use of the latest CMB measurements from final 2018 Planck legacy release \cite{Aghanim:2018eyx,Aghanim:2018oex,Aghanim:2019ame}.
    
    \item {\bf Baryon Acoustic Oscillations (BAO)}: a number of BAO constraints from different astronomical missions are exploited, namely those from 6dFGS~\cite{Beutler:2011hx}, SDSS-MGS~\cite{Ross:2014qpa}, and BOSS DR12~\cite{Alam:2016hwk} surveys, as considered by the Planck collaboration~\cite{Aghanim:2018eyx}.

    \item {\bf R19}: we adopt the latest measurement of the Hubble constant, obtained by a reanalysis of the Hubble Space Telescope data using Cepheids as calibrators, i.e. $H_0 = 74.03 \pm 1.42$ km/s/Mpc at $68\%$ CL~\cite{Riess:2019cxk}.
    
    \item {\bf Pantheon}: we analyse the luminosity distance data of type Ia supernovae from the Pantheon catalog \cite{Scolnic:2017caz}, including 1048 data points in the redshift region $z \in [0.01, 2.3]$.
    
    \item {\bf Cosmic Chronometers (CC)}: thirty measurements of the Hubble parameter at different redshifts extracted from Cosmic Chronometers in the redshift range $0< z < 2$ (as tabulated in \cite{Moresco:2016mzx}) are also considered in our data analyses.  
    
\end{itemize}

We consider a fiducial cosmology described by eight cosmological parameters: six of the standard $\Lambda$CDM model (the baryon and the cold dark matter energy densities $\Omega_{\rm b}h^2$ and $\Omega_{\rm c}h^2$, the ratio between the sound horizon and the 
angular diameter distance at decoupling $100\theta_{MC}$, the reionization optical depth 
$\tau$, and the spectral index and the amplitude of the scalar primordial power spectrum 
$n_{s}$ and $A_{s}$) and two accounting for the dark sector physics (the dark energy equation of state $w_{x}$ and the strength of the coupling $\xi$). Therefore, the Interacting Dark Energy (IDE) scenario is described by: 

\begin{eqnarray}
\mathcal{P} \equiv\Bigl\{\Omega_{b}h^2, \Omega_{c}h^2, 100\theta_{MC}, \tau, n_{s}, log[10^{10}A_{s}], \xi, w_{x} \Bigr\}~.
\label{eq:parameter_space1}
\end{eqnarray}

We consider $\xi<0$ in the phantom scenario ($w_{x}<-1$) and $\xi>0$ in the quintessence regime ($w_{x}>-1$) to avoid early-time instabilities~\cite{Valiviita:2008iv,Gavela:2009cy}, see Tab.~\ref{priors} for the priors on all the parameters. It is important to mention that the division of the parameter space into ($\xi < 0$, $w_x < -1$) or ($\xi > 0$, $w_x > -1$) is motivated from the doom factor analysis \cite{Gavela:2009cy}. Following the notation of \cite{Yang:2017zjs,Yang:2017ccc,Yang:2018euj}, the doom factor $d$ is defined as, $d = - Q\; [3H (1+w_x) \rho_x]^{-1}$. The interacting model becomes stable for $d < 0$ which for the present model $Q = 3 H \xi \rho_x$ requires either $\xi < 0$, $w_x < -1$ or $\xi > 0$, $w_x > -1$.  Hence, as we can see this division is necessary to maintain the early time instability in the perturbation evolution of the scalar modes of the dark species (DM and DE).  
In this manuscript, we shall refer to the scenario with a phantom-like DE equation of state ($w_x < -1$) as IDEp and by IDEq to the IDE scenario with a quintessence-like DE equation of state ($w_x > -1$). 

Then we allow for freedom in the dark radiation sector, by enlarging the fiducial cosmological scenario with the sum of the neutrino masses $M_{\nu}$ (IDE $+$ $M_{\nu}$): 

\begin{eqnarray}
\mathcal{P} \equiv\Bigl\{\Omega_{b}h^2, \Omega_{c}h^2, 100\theta_{MC}, \tau, n_{s}, \log[10^{10}A_{s}], 
\xi, w_x, M_{\nu}\Bigr\}~,
\label{eq:parameter_space2}
\end{eqnarray}

\noindent or with extra relativistic species at recombination  (IDE $+$ $N_{\rm eff}$): 

\begin{eqnarray}
\mathcal{P} \equiv\Bigl\{\Omega_{b}h^2, \Omega_{c}h^2, 100\theta_{MC}, \tau, n_{s}, \log[10^{10}A_{s}], \nonumber\\ 
\xi, w_x, N_{\rm eff} \Bigr\}~.
\label{eq:parameter_space3}
\end{eqnarray}

Finally, we also analyze the full scenario IDE $+$ $M_{\nu}$ $+$ $N_{\rm eff}$:
\begin{eqnarray}
\mathcal{P} \equiv\Bigl\{\Omega_{b}h^2, \Omega_{c}h^2, 100\theta_{MC}, \tau, n_{s}, \log[10^{10}A_{s}], \nonumber\\ 
\xi, w_x, M_{\nu}, N_{\rm eff} \Bigr\}~.
\label{eq:parameter_space3}
\end{eqnarray}

\begin{table}
\begin{center}
\begin{tabular}{|c|c|c|}
\hline
Parameter                    & prior phantom & prior quintessence\\
\hline
$\Omega_{\rm b} h^2$         & $[0.013,0.033]$& $[0.013,0.033]$ \\
$\Omega_{\rm c} h^2$         & $[0.001,0.99]$& $[0.001,0.99]$ \\
$100\theta_{MC}$             & $[0.5,10]$& $[0.5,10]$ \\
$\tau$                       & $[0.01,0.8]$& $[0.01,0.8]$ \\
$n_\mathrm{S}$               & $[0.7,1.3]$& $[0.7,1.3]$ \\
$\log[10^{10}A_{s}]$                       & $[1.7, 5.0]$& $[1.7, 5.0]$ \\
$w_x$                        & $[-3,-1]$&  $[-1,0]$  \\
$\xi$                        & $[-1,0]$& $ [0,1]$  \\
$M_{\nu}$                    & $[0,1]$ & $[0,1]$ \\
$N_{\rm eff}$                & $[0.05,10]$& $[0.05,10]$  \\
\hline %S
\end{tabular}
\end{center}
\caption{List of the flat priors on the cosmological parameters assumed in 
this work.}
\label{priors}
\end{table}

For numerical purposes, we make use of the latest version of the publicly available Markov Chain Monte Carlo code \texttt{CosmoMC}~\cite{Lewis:2002ah,Lewis:1999bs} package which supports the new 2018 Planck likelihood~\cite{Aghanim:2019ame} and that has been modified to include IDE scenarios. The  convergence diagnostic follows the Gelman-Rubin criteria~\cite{Gelman-Rubin}.

\section{Results}
\label{sec-4}
Throughout this section we will present the results obtained within the different IDE scenarios.

\subsection{IDE with $w_x<-1$}
Here we will start analysing the constraints obtained under the assumption of a phantom-like dark energy equation of state $w_x<-1$. 

\subsubsection{IDE}
\begin{figure*}
\includegraphics[width=0.6\textwidth]{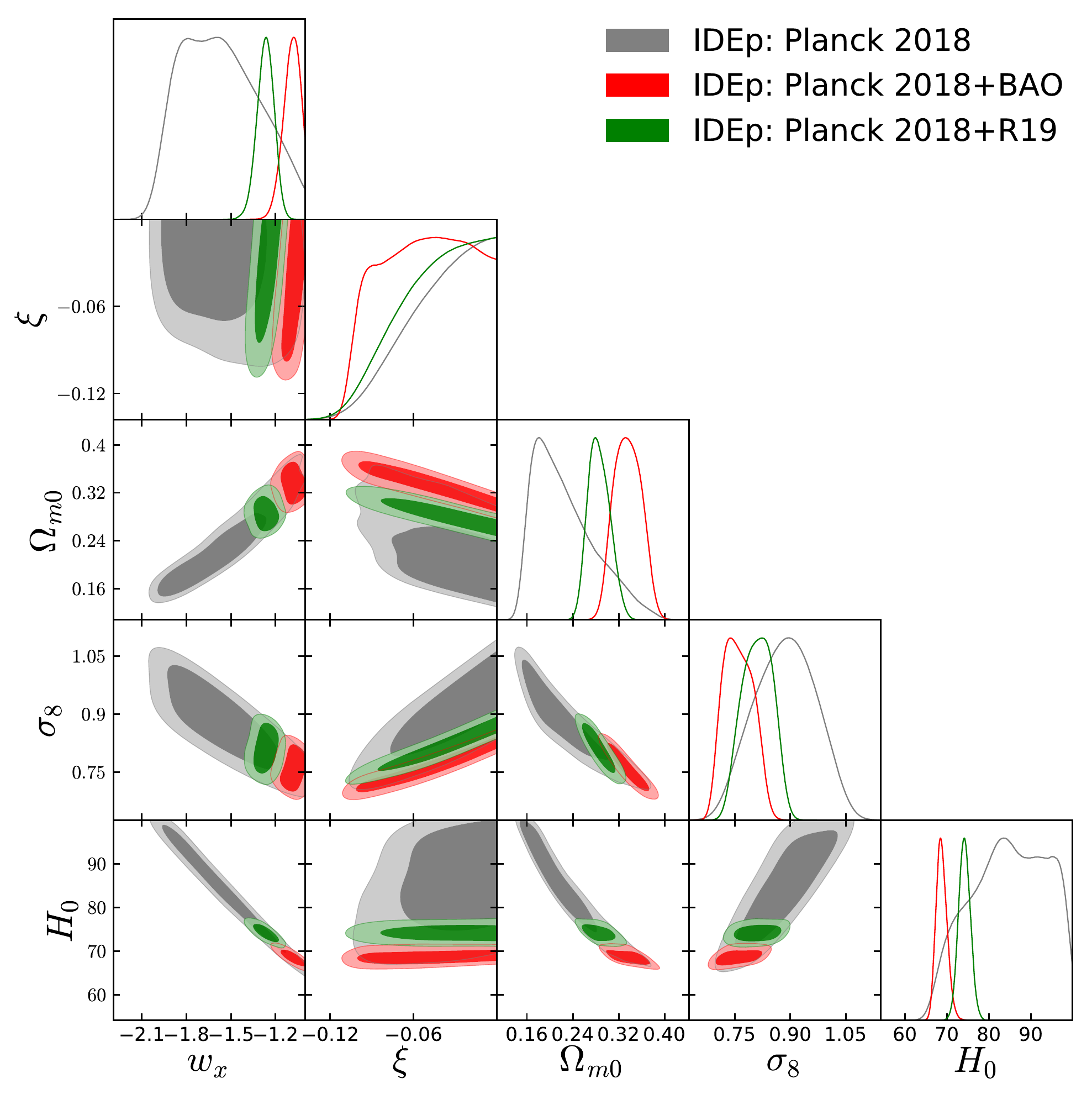}
\caption{68\% and 95\%~CL allowed contours and one-dimensional posterior probability distributions within the minimal IDE model in the phantom-like dark energy scenario ($w_x< -1$) from several data set combinations. Only a sub-set of cosmological parameters are shown.}
\label{fig:figures-IDEp}
\end{figure*}

The results for our baseline IDE scenario, based on eight parameters, with a dark energy equation of state $w_x<-1$ are shown in Tab.~\ref{tab:results-IDEp} and Fig.~\ref{fig:figures-IDEp}. 

Notice that, regardless of the dataset combination, the CDM energy density $\Omega_ch^2$ is larger than within the $\Lambda$CDM model. This is mainly due to the energy flow between the dark sectors, which, for this phantom case with $\xi <0$ (i.e. energy flux flowing from DE to DM) results in a larger density for cold dark matter at present, see also~\cite{DiValentino:2019jae}. Also, the Hubble constant is always much larger than in the  canonical $\Lambda$CDM scenario when considering CMB only, due to the fact that in the phantom region there is a strong degeneracy between $w_x$ and $H_0$ at the level of the CMB (see Fig.~\ref{fig:figures-IDEp}). When the dark energy equation of state is allowed to vary in the $w<-1$ region, $H_0$ must be larger to prevent a shift in the CMB peaks. Notice that here, there is no preference for a non-zero dark sector coupling, and the so-called Hubble constant tension is strongly alleviated due to the phantom character of the DE component and not from the presence of a coupling. From CMB measurements alone, we find that a dark energy equation of state $w_x<-1$ is preferred at a significance above the 2$\sigma$ level and a lower limit of $\xi>-0.090$ at 95\% CL. The value of $S_8$ is instead shifted in the right direction to solve the tension of Planck with the cosmic shear experiments DES~\cite{Abbott:2017wau,Troxel:2017xyo}, KiDS-450~\cite{Kuijken:2015vca,Hildebrandt:2016iqg,Conti:2016gav} or CFHTLenS~\cite{Heymans:2012gg, Erben:2012zw,Joudaki:2016mvz}, i.e.  $S_8=0.756\pm0.034$ at 68\% CL for CMB in the IDE model, to be compared with $S_8=0.822\pm0.015$ at 68\% CL in the $\Lambda$CDM scenario for the same data set, or $S_8=0.777^{+0.022}_{-0.036}$ at 68\% CL for a $w$CDM cosmology. When including the BAO data, see the third column of Tab.~\ref{tab:results-IDEp}, the six parameters of the standard $\Lambda$CDM model are almost unmodified, while $w_x$ is now perfectly consistent with a cosmological constant at the 2$\sigma$ level. For CMB+BAO, $H_0$ shifts back towards a  lower value ($H_0=68.7^{+2.7}_{-2.5}$ km/s/Mpc at 95\% CL). The value of the clustering parameter $\sigma_8$ is still below the $\Lambda$CDM and $w$CDM ones, due to the correlation with $\xi$, see Fig.~\ref{fig:figures-IDEp}. When considering the combination of CMB+R19, shown in the fourth column of Tab.~\ref{tab:results-IDEp}, we find a value for the Hubble constant in agreement with the R19 measurement, together with a preference at more than 3$\sigma$ for a phantom dark energy ($w_x=-1.27^{+0.14}_{-0.17}$ at 99\% CL). Adding Pantheon and CC measurements to the CMB (see the fifth column of Tab.~\ref{tab:results-IDEp}) leads to constraints very similar to the CMB+BAO case. The combination CMB+BAO+Pantheon+CC, given in the sixth column of Tab.~\ref{tab:results-IDEp}, only diminishes the error bars, when compared to the previous case. Finally, the data combination of CMB+R19+Pantheon+CC, see the last column of Tab.~\ref{tab:results-IDEp}, is completely driven by the discrepancies between the CMB+Pantheon+CC data combination and the R19 measurements. Therefore, the Hubble constant  shifts slightly towards higher values, reducing the tension with R19 at 2.3$\sigma$ and leading to an evidence for a phantom dark energy component $w_x<-1$ at a high significance.
\begingroup
\squeezetable
\begin{center}
\begin{table*}
\scalebox{1}{
\begin{tabular}{ccccccccc}
\hline\hline
Parameters & CMB & CB & CR19 & CPCC & CBPCC & CR19PCC\\ 
\hline
$\Omega_ch^2$&
$0.134_{-0.015}^{+0.017}$&
$0.135_{-0.014}^{+0.014}$&
$0.134_{-0.015}^{+0.017}$ & 
$    0.135_{-    0.014}^{+    0.015}$ & 
$    0.135_{-    0.014}^{+    0.014}$ & 
$    0.134_{-    0.015}^{+    0.016}$
\\

$\Omega_bh^2$&
$0.02239_{-0.00030}^{+0.00030}$&
$0.02239_{-0.00028}^{+0.00014+0.00029}$&
$0.02238_{-0.00029}^{+0.00029}$ &
$    0.02236_{-    0.00030}^{+    0.00030}$ &
$    0.02240_{-    0.00028}^{+    0.00028}$ &
$    0.02242_{-    0.00029}^{+    0.00028}$
\\

$100\theta_{MC}$&
$1.0402_{-0.0011}^{+0.0011}$&
$1.0401_{-0.0010}^{+0.0010}$&
$1.0402_{-0.0011}^{+0.0010}$ & 
$    1.0401_{-    0.0010}^{+    0.0010}$& 
$    1.0401_{-    0.0010}^{+    0.0010}$ &
$    1.0402_{-    0.0010}^{+    0.0010}$
\\

$\tau$&
$0.054_{-0.015}^{+0.015}$&
$0.055_{-0.015}^{+0.016}$&
$0.054_{-0.015}^{+0.016}$ &
$    0.054_{-    0.015}^{+    0.016}$ &
$    0.055_{-    0.015}^{+    0.016}$ &
$    0.055_{-    0.015}^{+    0.016}$
\\

$n_s$&
$0.9652_{-0.0086}^{+0.0086}$&
$0.9657_{-0.0082}^{+0.0081}$&
$0.9648_{-0.0086}^{+0.0086}$ &
$    0.9647_{-    0.0086}^{+        0.0086}$ &
$    0.9660_{-        0.0080}^{+    0.0078}$ &
$    0.9662_{-        0.0085}^{+        0.0085}$
\\

${\rm{ln}}(10^{10}A_s)$&
$3.043_{-0.031}^{+0.029}$&
$3.045_{-0.030}^{+0.032}$&
$3.044_{-0.030}^{+0.031}$ & 
$    3.045_{-        0.031}^{+       0.033}$ &
$    3.045_{-        0.031}^{+        0.033}$ &
$    3.045_{-        0.031}^{+        0.033}$
\\

$w_x$ &
$-1.58_{-0.44}^{+0.49}$&
$-1.094_{-0.099}^{+0.094}$&
$-1.27_{-0.12}^{+0.11}$ & 
$   -1.085_{-        0.078}^{+        0.083}$ &
$   -1.080_{-        0.072}^{+        0.078}$ &
$   -1.139_{-        0.083}^{+        0.081}$
\\

$\xi$&
$>-0.090$&
$>-0.101$&
$>-0.094$ & 
$>-0.100$ &
$>-0.101$ &
$>-0.099$ 
\\

$\Omega_{m0}$&
$0.23_{-0.09}^{+0.11}$&
$0.336_{-0.042}^{+0.044}$&
$0.286_{-0.035}^{+0.038}$ & 
$    0.338_{-        0.038}^{+        0.040}$ &
$    0.339_{-        0.036}^{+        0.037}$ &
$    0.318_{-       0.036}^{+        0.038}$
\\

$\sigma_8$&
$0.89_{-0.16}^{+0.15}$&
$0.760_{-0.070}^{+0.074}$&
$0.811_{-0.077}^{+0.072}$&
$    0.762_{-        0.063}^{+       0.066}$ &
 $    0.755_{-       0.064}^{+        0.068}$ &
 $    0.774_{-       0.066}^{+        0.066}$
\\

$H_0 {\rm [km/s/Mpc]}$&
$>68$&
$68.7_{-2.5}^{+2.7}$&
$74.2_{-2.7}^{+2.8}$ &
$   68.3_{-       1.9}^{+       2.0}$ &
$   68.3_{-      1.5}^{+        1.6}$ &
$   70.3_{-      1.7}^{+        1.7}$

\\

$S_8$&
$0.756_{-0.063}^{+0.064}$&
$0.802_{-0.036}^{+0.036}$&
$0.790_{-0.041}^{+0.040}$ & 
$    0.808_{-        0.039}^{+        0.039}$ &
$    0.801_{-        0.035}^{+        0.037}$ &
$    0.796_{-        0.040}^{+        0.040}$
\\

\hline\hline
\end{tabular}
}
\caption{$95\%$~CL constraints within the minimal IDE model in the phantom-like dark energy scenario ($w_x< -1$) from several dataset combinations, with CB, CR19, CPCC, CBPCC and CR19PCC referring to the combinations of CMB+BAO, CMB+R19, CMB+Pantheon+CC, CMB+BAO+Pantheon+CC and CMB+R19+Pantheon+CC, respectively.}
\label{tab:results-IDEp}
\end{table*}
\end{center}
\endgroup

\subsubsection{IDE + $M_\nu$ - 9 parameters}

\begin{figure*}
\includegraphics[width=0.6\textwidth]{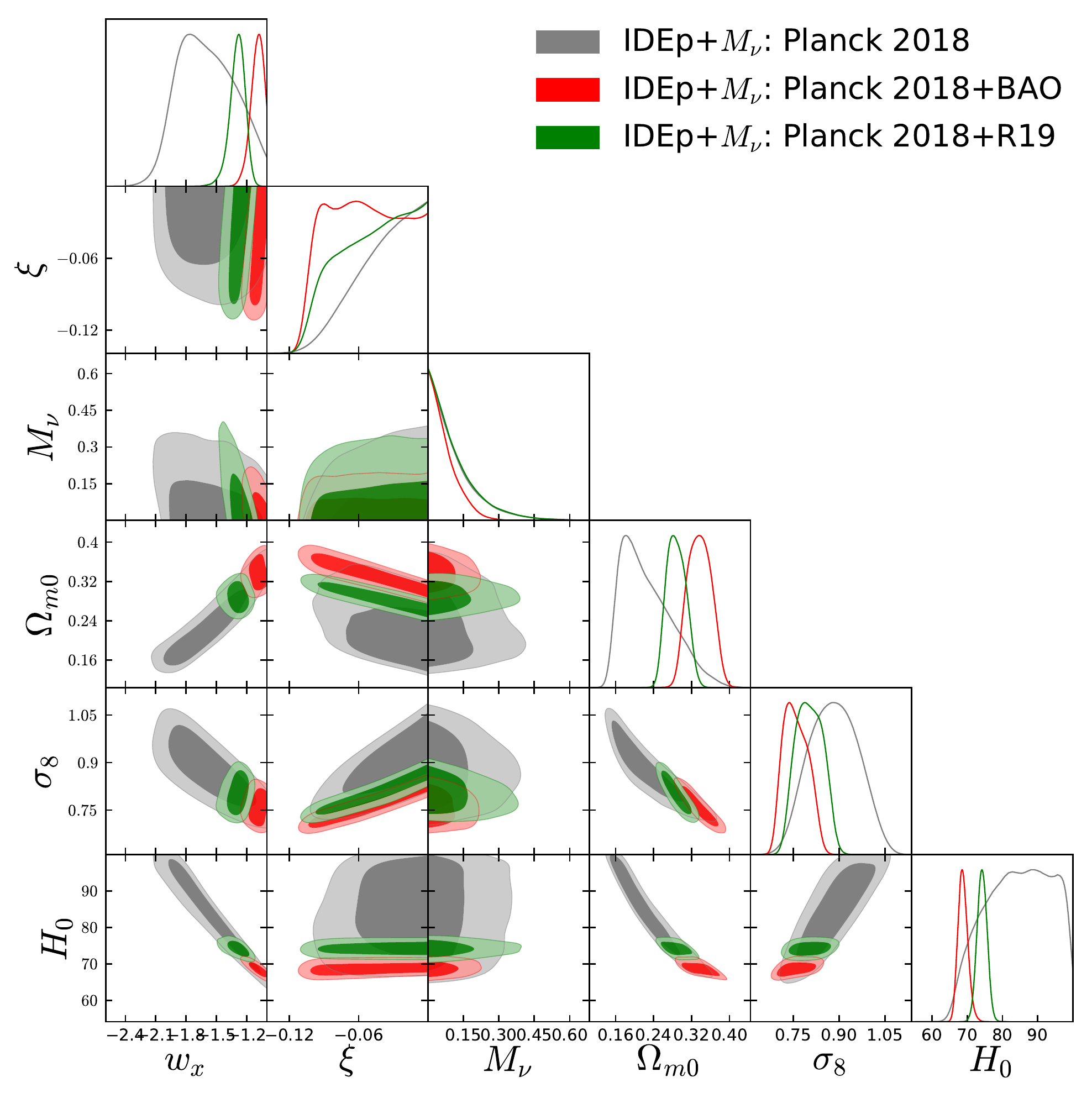}
\caption{68\% and 95\%~CL allowed contours and one-dimensional posterior probability distributions within the IDE$+M_{\nu}$ model in the phantom-like dark energy scenario ($w_x< -1$) from several data set combinations. Only a sub-set of cosmological parameters are shown.}
\label{fig:figures-IDEpmnu}
\end{figure*}

The results for an IDE + $M_\nu$ scenario, based on nine parameters, with a dark energy equation of state $w_x<-1$ are shown in Tab.~\ref{tab:results-IDEpmnu} and Fig.~\ref{fig:figures-IDEpmnu}.

In this scenario, extended to include the total neutrino mass $M_\nu$, independently from the combination of data chosen, all the constraints on the six cosmological parameters of the $\Lambda$CDM model, as well as on the dark energy equation of state $w_x$ and on all the derived parameters $H_0$, $\Omega_{m0}$, $\sigma_8$ and $S_8$, are very similar to the minimal IDE scenario of the previous section (see Tab.~\ref{tab:results-IDEp}). Therefore, the addition of the total neutrino mass is not introducing new correlations between the previous cosmological parameters, as we can see in Fig.~\ref{fig:figures-IDEpmnu}. The well-known correlation between $M_\nu$ and the dark energy equation of state $w_x$, the matter density $\Omega_{m0}$ and the clustering parameter $\sigma_8$ are instead present also in this model, especially visible for the CMB+BAO and CMB+R19 combination of data. However, the also very well-known degeneracy between $H_0$ and $M_\nu$ disappears completely in this scenario. For this reason, the $M_\nu$ and $\Omega_\nu h^2$ upper limits are identical in the CMB and CMB+R19 cases, instead of improving  when R19 measurements are added to CMB observations (as it happens within the $\Lambda$CDM + $M_\nu$ model). Comparing the constraints obtained in this extended IDE scenario with those obtained within a $\Lambda$CDM scenario for the same combination of data sets, one can notice that the upper limits on $M_\nu$ are slightly relaxed in this case, because of the phantom behavior of the dark energy equation of state and the $M_\nu$ -- $w_x$ correlation. Cosmological neutrino mass bounds are softened if the dark energy equation of state is taken as a free parameter. If $w_x$ is allowed to vary, the matter energy density takes very high values, and that can be compensated with a larger neutrino mass. Indeed, the authors of Ref.~\cite{LaVacca:2008mh} found that a larger neutrino mass would be allowed when a dark coupling is present. The most stringent limit we find on the total neutrino mass is for the  CMB+R19+Pantheon+CC data combination, $M_\nu<0.151$ eV at 95\% CL. The former bound is very similar to the one obtained for CMB+BAO+Pantheon+CC, $M_\nu<0.156$ eV at 95\% CL.

\begingroup
\squeezetable
\begin{center}
\begin{table*}
\scalebox{1}{
\begin{tabular}{cccccccccccccc}
\hline\hline
Parameters & CMB & CB & CR19 & CPCC & CBPCC & CR19PCC \\ \hline
$\Omega_ch^2$&
$0.133_{-0.015}^{+0.018}$&
$0.136_{-0.016}^{+0.015}$&
$0.135_{-0.015}^{+0.017}$&
$    0.135_{-       0.014}^{+     0.014}$ &
$    0.135_{-     0.015}^{+       0.015}$ &
$    0.134_{-     0.015}^{+       0.016}$
\\

$\Omega_bh^2$&
$0.02237_{-0.00031}^{+0.00030}$&
$0.02238_{-0.00028}^{+0.00028}$&
$0.02236_{-0.00031}^{+0.00031}$ &
$    0.02235_{-      0.00029}^{+        0.00030}$ &
$    0.02240_{-       0.00027}^{+        0.00028}$&
$    0.02243_{-       0.00029}^{+        0.00029}$
\\

$100\theta_{MC}$&
$1.0402_{-0.0011}^{+0.0010}$&
$1.0401_{-0.0010}^{+0.0010}$&
$1.0401_{-0.0010}^{+0.0010}$&
$    1.0401_{-        0.0010}^{+        0.0010}$ &
$    1.0401_{-        0.0010}^{+       0.0010}$ &
$    1.0402_{-        0.0011}^{+        0.0011}$
\\

$\tau$&
$0.054_{-0.015}^{+0.016}$&
$0.055_{-0.015}^{+0.016}$&
$0.054_{-0.015}^{+0.016}$ &
 $    0.054_{-      0.015}^{+    0.016}$ &
 $    0.055_{-      0.015}^{+        0.016}$ &
 $    0.055_{-      0.015}^{+        0.017}$
\\

$n_s$&
$0.9650_{-0.0088}^{+0.0088}$&
$0.9654_{-0.0078}^{+0.0080}$&
$0.9646_{-0.0091}^{+0.0087}$ &
$    0.9645_{-        0.0086}^{+        0.0086}$ &
$    0.9660_{-        0.0079}^{+        0.0080}$ &
$    0.9663_{-        0.0085}^{+        0.0085}$
\\

${\rm{ln}}(10^{10}A_s)$&
$3.043_{-0.031}^{+0.033}$&
$3.045_{-0.031}^{+0.034}$&
$3.044_{-0.032}^{+0.033}$ & 
$    3.045_{-        0.031}^{+    0.032}$ &
$    3.045_{-        0.031}^{+        0.033}$&
$    3.045_{-        0.031}^{+        0.035}$

\\

$w_x$&
$-1.60_{-0.46}^{+0.52}$&
$>-1.22$&
$-1.29_{-0.16}^{+0.15}$ &
$   -1.092_{-       0.086}^{+    0.090}$&
$ >-1.15$ &
$   -1.136_{-       0.087}^{+      0.082}$ 
\\

$\xi$&
$>-0.086$&
$>-0.101$&
$>-0.098$ & 
$ >-0.102$ &
$ >-0.101$ &
$>-0.100$
\\

$\Omega_{m0}$&
$0.23_{-0.09}^{+0.11}$&
$0.337_{0.043}^{+0.045}$&
$0.288_{-0.037}^{+0.039}$ & 
$    0.342_{-        0.042}^{+    0.042}$ &
$    0.339_{-        0.037}^{+        0.037}$ &
$    0.318_{-       0.038}^{+        0.039}$
\\

$\sigma_8$&
$0.88_{-0.14}^{+0.15}$&
$0.758_{-0.070}^{+0.076}$&
$0.803_{-0.076}^{+0.079}$ &
$    0.756_{-        0.065}^{+       0.071}$ &
$    0.756_{-    0.065}^{+       0.070}$ &
$    0.775_{-    0.069}^{+      0.070}$ 
\\

$H_0{\rm[km/s/Mpc]}$&
$>68$&
$68.8_{-2.7}^{+2.8}$&
$74.2_{-2.7}^{+2.7}$ & 
$   68.2_{-    2.1}^{+      2.1}$ &
$   68.3_{- 1.5}^{+    1.5}$ &
$   70.4_{-      1.7}^{+    1.7}$
\\

$M_{\nu}{\rm[eV]}$& 
$ < 0.313 $&
$< 0.183 $ &
$< 0.313$ & 
$<0.255$ &
 $<0.156$ &
 $<0.151$
\\

$\Omega_\nu h^2$&
$<0.0034$&
$<0.0020$&
$<0.0034$ &
$<0.0027$ &
$<0.0017$ &
$<0.0016$
\\

$S_8$&
$0.754_{-0.064}^{+0.069}$&
$0.802_{-0.037}^{+0.039}$&
$0.785_{-0.045}^{+0.044}$ & 
$    0.806_{-       0.041}^{+       0.042}$  & 
$    0.802_{-        0.037}^{+      0.038}$ &
$    0.796_{-        0.040}^{+       0.042}$
\\

\hline\hline
\end{tabular}
}
\caption{$95\%$~CL constraints within the IDE+$M_{\nu}$ model in the phantom-like dark energy scenario ($w_x< -1$) from several dataset combinations, with CB, CR19, CPCC, CBPCC and CR19PCC referring to the combinations of CMB+BAO, CMB+R19, CMB+Pantheon+CC, CMB+BAO+Pantheon+CC and CMB+R19+Pantheon+CC, respectively.}
\label{tab:results-IDEpmnu}
\end{table*}
\end{center}                                                \endgroup

\subsubsection{IDE + $N_{\rm eff}$ - 9 parameters}
\begin{figure*}
\includegraphics[width=0.6\textwidth]{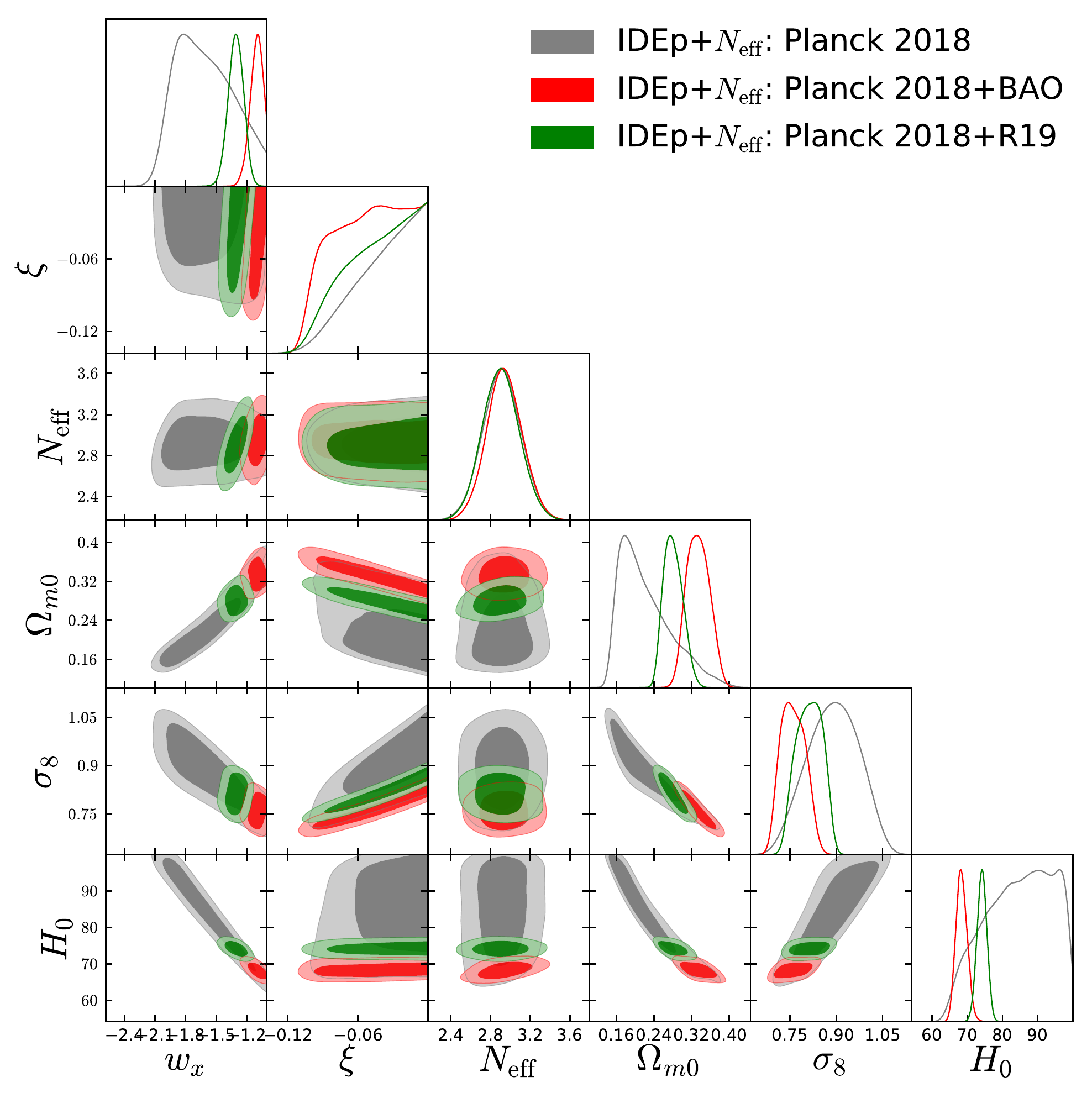}
\caption{68\% and 95\%~CL allowed contours and one-dimensional posterior probability distributions within the IDE$+N_{\rm eff}$ model in the phantom-like dark energy scenario ($w_x< -1$) from several data set combinations. Only a sub-set of cosmological parameters are shown.}
\label{fig:figures-IDEpneff}
\end{figure*}

The results for an IDE + $N_{\rm eff}$ scenario, based on nine parameters, with a dark energy equation of state $w_x<-1$ are shown in Tab.~\ref{tab:results-IDEpneff} and Fig.~\ref{fig:figures-IDEpneff}. Notice that in general the constraints on the cosmological parameters of the $\Lambda$CDM model, with the exception of the spectral index $n_s$ are very similar to those obtained in previous scenarios. Due to the well-known correlation between $N_{\rm eff}$ and $n_s$, we notice a very mild shift in the mean value of the spectral index toward smaller values. The effective number of relativistic degrees of freedom $N_{\rm eff}$ we recover in this modified cosmological scenario is perfectly consistent with its standard value $N_{\rm eff}=3.046$~\cite{Mangano:2005cc,deSalas:2016ztq}.
Also in this scenario, the addition of the neutrino effective number $N_{\rm eff}$ as a free parameter is not introducing a new direction of correlation between the previous cosmological parameters, see Fig.~\ref{fig:figures-IDEpneff}. 
The well-known correlation between $N_{\rm eff}$ and the Hubble constant $H_0$ disappears completely in this scenario, since the dark energy equation of state is also an additional free parameter. For this very same reason, the  constraints on $N_{\rm eff}$ are the same for the CMB and CMB+R19 cases. When adding Pantheon and CC data to the CMB (see the fifth column of the Tab.~\ref{tab:results-IDEpneff}), the relativistic degrees of freedom shift towards higher values ($N_{\rm eff}=2.99\pm 0.16$ at 68\% CL) with respect to the CMB only case (where $N_{\rm eff}=2.91\pm 0.019$ at 68\% CL). When we further add the R19 measurement, i.e. we consider CMB+R19+Pantheon+CC, see the last column of Tab.~\ref{tab:results-IDEpneff}, we obtain our most stringent bound $N_{\rm eff}=3.22^{+0.31}_{-0.29}$ at 95\% CL associated to the highest mean value for $N_{\rm eff}$.

\begingroup
\squeezetable
\begin{center}
\begin{table*}
\scalebox{1}{
\begin{tabular}{ccccccccccccccccc}
\hline\hline
Parameters & CMB & CB & CR19 & CPCC & CBPCC & CR19PCC \\ \hline
$\Omega_ch^2$&
$0.131_{-0.017}^{+0.019}$&
$0.133_{-0.017}^{+0.017}$&
$0.131_{-0.016}^{+0.018}$ &
$    0.134_{-        0.017}^{+        0.017}$ &
$    0.135_{-        0.017}^{+        0.017}$ &
$    0.138_{-        0.017}^{+        0.018}$
\\

$\Omega_bh^2$&
$0.02227_{-0.00043}^{+0.00042}$&
$0.02230_{-0.00039}^{+0.00039}$&
$0.02226_{-0.00043}^{+0.00043}$ &
$    0.02232_{-        0.00039}^{+       0.00038}$ &
$    0.02238_{-        0.00035}^{+        0.00036}$ &
$    0.02255_{-       0.00036}^{+       0.00037}$
\\

$100\theta_{MC}$&
$1.0404_{-0.0013}^{+0.0012}$&
$1.0404_{-0.0013}^{+0.0012}$&
$1.0404_{-0.0013}^{+0.0012}$ &
$    1.0402_{-       0.0012}^{+      0.0012}$ &
$    1.0402_{-        0.0012}^{+       0.0012}$ &
$    1.0399_{-       0.0012}^{+      0.0012}$
\\

$\tau$&
$0.053_{-0.016}^{+0.016}$&
$0.054_{-0.015}^{+0.016}$&
$0.053_{-0.015}^{+0.016}$ &
$    0.054_{-        0.015}^{+      0.016}$ &
$    0.055_{-    0.015}^{+        0.016}$ &
$    0.056_{-        0.015}^{+     0.017}$
\\

$n_s$&
$0.960_{-0.017}^{+0.017}$&
$0.962_{-0.015}^{+0.015}$&
$0.960_{-0.017}^{+0.017}$ &
$    0.963_{-      0.015}^{+        0.015}$  &
$    0.965_{-     0.014}^{+        0.014}$ &
$    0.972_{-     0.013}^{+        0.013}$
\\

${\rm{ln}}(10^{10}A_s)$&
$3.037_{-0.038}^{+0.037}$&
$3.039_{-0.036}^{+0.036}$&
$3.036_{-0.037}^{+0.037}$ &
$    3.042_{-    0.036}^{+       0.036}$ &
$    3.043_{-    0.035}^{+       0.036}$ &
$    3.053_{-    0.034}^{+       0.036}$
\\

$w_x$&
$-1.61_{-0.46}^{+0.54}$&
$>-1.22$&
$-1.31_{-0.15}^{+0.14}$ &
$   -1.090_{-        0.079}^{+        0.087}$ &
$   -1.082_{-        0.074}^{+        0.081}$ &
$   -1.122_{-        0.088}^{+        0.084}$
\\

$\xi$&
$>-0.087$&
$>-0.099$&
$>-0.094$ &
$ >-0.101$&
$ >-0.101$ &
$ >-0.101$
\\

$\Omega_{m0}$&
$0.22_{-0.09}^{+0.12}$&
$0.334_{-0.042}^{+0.046}$&
$0.281_{-0.037}^{+0.039}$ &
$    0.340_{--    0.041}^{+        0.040}$ &
$    0.339_{-        0.036}^{+       0.036}$ &
$    0.321_{-        0.036}^{+       0.037}$
\\

$\sigma_8$&
$0.89_{-0.16}^{+0.15}$&
$0.761_{-0.071}^{+0.074}$&
$0.814_{-0.075}^{+0.072}$ &
$    0.758_{-        0.064}^{+        0.068}$&
$    0.754_{-        0.064}^{+        0.068}$ &
$    0.773_{-        0.069}^{+        0.070}$
\\

$H_0{\rm[km/s/Mpc]}$&
$>67$&
$68.4_{-2.9}^{+3.2}$&
$74.2_{-2.8}^{+2.7}$ &
$   68.0_{-       2.5}^{+       2.6}$ &
$   68.2_{-       2.1}^{+       2.1}$ &
$   70.9_{-       2.0}^{+       2.0}$
\\

$N_{\rm eff}$&
$2.91_{-0.38}^{+0.38}$&
$2.94_{-0.35}^{+0.36}$&
$2.90_{-0.36}^{+0.38}$ &
$    2.99_{-        0.31}^{+        0.32}$ &
$    3.02_{-        0.31}^{+        0.32}$  &
$    3.22_{-        0.29}^{+        0.31}$
\\

$S_8$&
$0.753_{-0.065}^{+0.068}$&
$0.801_{-0.035}^{+0.036}$&
$0.785_{-0.041}^{+0.041}$ &
$    0.806_{-    0.040}^{+        0.040}$ &
$    0.800_{-        0.035}^{+        0.037}$ &
$    0.798_{-        0.040}^{+        0.041}$
\\
\hline\hline
\end{tabular}
}
\caption{$95\%$~CL constraints within the IDE$+N_{\rm eff}$ scheme in the phantom-like dark energy scenario ($w_x< -1$) from several dataset combinations, with CB, CR19, CPCC, CBPCC and CR19PCC referring to the combinations of CMB+BAO, CMB+R19, CMB+Pantheon+CC, CMB+BAO+Pantheon+CC and CMB+R19+Pantheon+CC, respectively.}
\label{tab:results-IDEpneff}
\end{table*}
\end{center}
\endgroup

\subsubsection{IDE + $M_\nu$ + $N_{\rm eff}$ - 10 parameters}

\begin{figure*}
\includegraphics[width=0.6\textwidth]{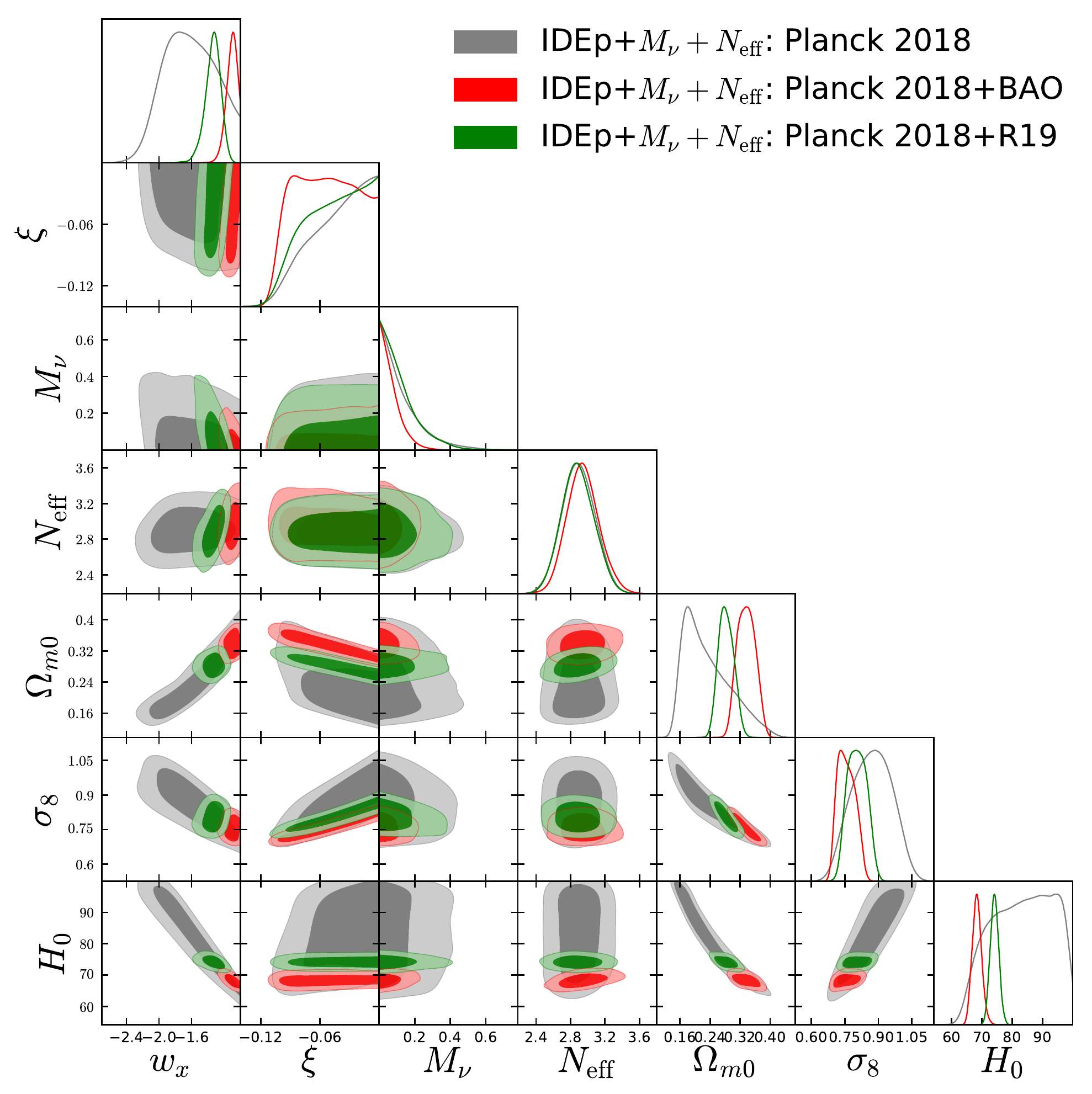}
\caption{68\% and 95\%~CL allowed contours and one-dimensional posterior probability distributions within the IDE$+M_{\nu}+N_{\rm eff}$ model in the phantom-like dark energy scenario ($w_x< -1$) from several data set combinations. Only a sub-set of cosmological parameters are shown.}
\label{fig:figures-IDEpmnuneff}
\end{figure*}

The results for an IDE + $M_\nu$ + $N_{\rm eff}$ scenario, based on ten parameters, with a dark energy equation of state $w_x<-1$ are shown in Tab.~\ref{tab:results-IDEpmnuneff} and Fig.~\ref{fig:figures-IDEpmnuneff}. In this scenario the spectral index $n_s$ is shifting further down and we obtain an indication for $w_x$ to be phantom at a high significance level for the CMB+R19 data set combination, see Tab.~\ref{tab:results-IDEpmnuneff}.

The considerations we made in the previous cases about the directions of the correlations between the cosmological parameters are unaltered by the variation of $M_\nu$ and $N_{\rm eff}$ at the same time. While the constraints on the neutrino effective number $N_{\rm eff}$ for all the data set combinations are identical to the IDE+$N_{\rm eff}$ scenario,  the upper limits on the total neutrino mass $M_\nu$ are mildly relaxed with respect to the IDE+$M_\nu$ model. The most stringent bound on this parameter is now obtained for the combination CMB+BAO+Pantheon+CC (see the sixth column of Tab.~\ref{tab:results-IDEpmnuneff}), and it corresponds to $M_\nu<0.160$~eV at 95\% CL.

\begingroup
\squeezetable
\begin{center}
\begin{table*}
\scalebox{1}{
\begin{tabular}{ccccccccccccccccc}
\hline\hline
Parameters & CMB & CB & CR19 & CPCC & CBPCC & CR19PCC \\ \hline
$\Omega_ch^2$&
$0.132_{-0.017}^{+0.019}$&
$0.134_{-0.018}^{+0.017}$&
$0.132_{-0.017}^{+0.018}$ &
$    0.135_{-        0.017}^{+        0.017}$ &
$    0.135_{-        0.017}^{+        0.017}$ &
$    0.139_{-        0.018}^{+        0.018}$
\\

$\Omega_bh^2$&
$0.02223_{-0.00045}^{+0.00046}$&
$0.02230_{-0.00040}^{+0.00042}$&
$0.02223_{-0.00044}^{+0.00043}$ &
$    0.02232_{-        0.00038}^{+        0.00038}$ &
$    0.02237_{-        0.00036}^{+        0.00036}$ &
$    0.02256_{-        0.00035}^{+        0.00036}$
\\

$100\theta_{MC}$&
$1.0404_{-0.0013}^{+0.0013}$&
$1.0403_{-0.0012}^{+0.0013}$&
$1.0404_{-0.0012}^{+0.0012}$ & 
$    1.0402_{-      0.0012}^{+        0.0012}$ &
$    1.0402_{-        0.0012}^{+        0.0012}$ &
$    1.0399_{-        0.0012}^{+        0.0012}$
\\

$\tau$&
$0.053_{-0.016}^{+0.016}$&
$0.054_{-0.015}^{+0.016}$&
$0.053_{-0.015}^{+0.017}$ &
$    0.054_{-        0.015}^{+        0.016}$ &
$    0.055_{-        0.015}^{+        0.017}$ &
$    0.056_{-        0.016}^{+        0.017}$
\\

$n_s$&
$0.959_{-0.017}^{+0.017}$&
$0.962_{-0.015}^{+0.016}$&
$0.959_{-0.016}^{+0.017}$ &
$    0.963_{-       0.015}^{+        0.015}$ &
 $    0.965_{-       0.014}^{+       0.014}$ & 
 $    0.973_{-        0.013}^{+        0.013}$
\\

${\rm{ln}}(10^{10}A_s)$&
$3.036_{-0.037}^{+0.037}$&
$3.040_{-0.036}^{+0.037}$&
$3.037_{-0.037}^{+0.039}$ &
$    3.043_{-        0.034}^{+        0.036}$  &
$    3.043_{-        0.034}^{+        0.037}$ &
$    3.054_{-        0.035}^{+        0.037}$
\\

$w_x$&
$>-2.13$&
$>-1.23$&
$-1.34_{0.20}^{+0.19}$ &
$   -1.097_{-        0.088}^{+       0.091}$ &
$>-1.158$ &
$   -1.123_{-       0.093}^{+        0.089}$
\\

$\xi$&
$>-0.093$&
$>-0.101$&
$>-0.097$ &
$ >-0.101$  &
$>-0.101$ &
$>-0.102$
\\

$\Omega_{m0}$&
$0.23_{-0.10}^{+0.13}$&
$0.336_{-0.042}^{+0.044}$&
$0.283_{-0.037}^{+0.040}$ &
$    0.342_{-        0.040}^{+        0.041}$ &
$    0.339_{-        0.037}^{+        0.037}$ &
$    0.322_{-        0.037}^{+        0.038}$ 
\\

$\sigma_8$&
$0.87_{-0.16}^{+0.16}$&
$0.757_{-0.068}^{+0.074}$&
$0.805_{-0.075}^{+0.079}$ &
$    0.755_{-        0.064}^{+        0.069}$  &
$    0.756_{-        0.065}^{+        0.071}$ &
$    0.772_{-        0.069}^{+        0.075}$
\\

$H_0{\rm[km/s/Mpc]}$&
$>65$&
$68.4_{-3.1}^{+3.1}$&
$74.2_{-2.7}^{+2.8}$ &
$   68.0_{-        2.7}^{+        2.7}$ &
$   68.2_{-        2.1}^{+        2.2}$ & 
$   70.9_{-       2.0}^{+      1.9}$
\\

$M_{\nu}{\rm[eV]} $&
$< 0.367$&
$< 0.184$&
$< 0.332$ &
$<0.263$ &
$<0.160$ & 
$<0.172$
\\

$N_{\rm eff}$&
$2.89_{-0.37}^{+0.37}$&
$2.94_{-0.35}^{+0.39}$&
$2.88_{-0.36}^{+0.37}$ &
$    3.00_{-        0.33}^{+        0.34}$ & 
$    3.02_{-        0.31}^{+        0.33}$ & 
$    3.22_{-        0.29}^{+        0.29}$
\\

$\Omega_\nu h^2$&
$<0.0037$&
$<0.0019$&
$<0.0034$ &
$<0.0028$ &
$<0.0017$ & 
$<0.0018$ 
\\

$S_8$&
$0.751_{-0.067}^{+0.068}$&
$0.800_{-0.037}^{+0.039}$&
$0.779_{-0.045}^{+0.044}$ & 
$    0.805_{-        0.040}^{+     0.041}$  &
$    0.803_{-        0.037}^{+        0.039}$  & 
$    0.798_{-        0.041}^{+        0.042}$

\\
\hline\hline
\end{tabular}
}
\caption{$95\%$~CL constraints within the IDE$+M_{\nu}+N_{\rm eff}$ scheme in the phantom-like dark energy scenario ($w_x< -1$) from several dataset combinations, with CB, CR19, CPCC, CBPCC and CR19PCC referring to the combinations of CMB+BAO, CMB+R19, CMB+Pantheon+CC, CMB+BAO+Pantheon+CC and CMB+R19+Pantheon+CC, respectively.}
\label{tab:results-IDEpmnuneff}
\end{table*}
\end{center}
\endgroup

\subsection{IDE with $w_x>-1$}
In the following sections, we shall present the constraints for the case of a quintessence-like dark energy equation of state $w_x>-1$.

\subsubsection{IDE}
\begin{figure*}
\includegraphics[width=0.6\textwidth]{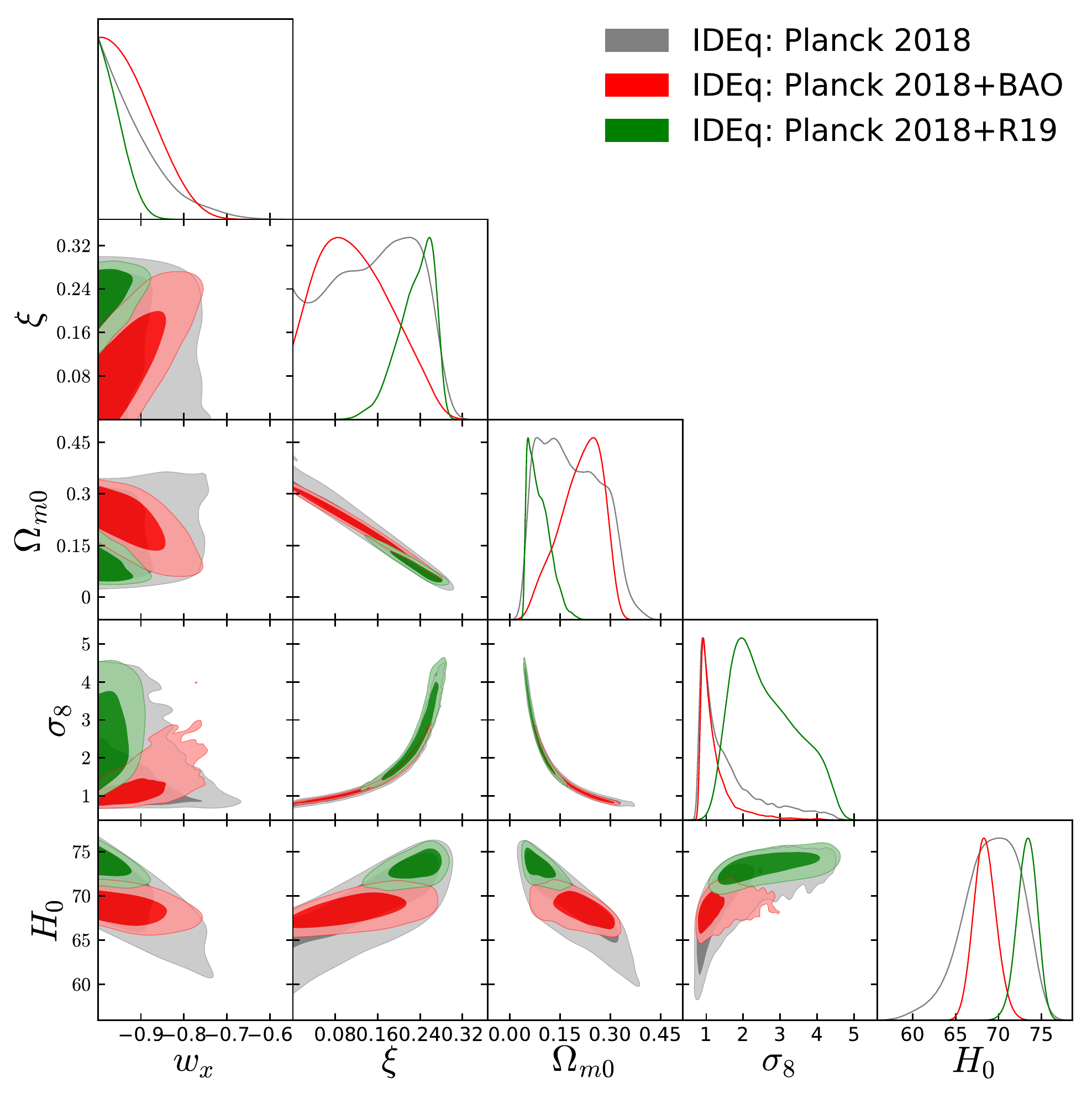}
\caption{68\% and 95\%~CL allowed contours and one-dimensional posterior probability distributions within the IDE model in the quintessence-like dark energy scenario ($w_x> -1$) from several data set combinations. Only a sub-set of cosmological parameters are shown.}
\label{fig:figures-IDEq}
\end{figure*}

The results for our baseline IDE scenario, based on eight parameters, with a dark energy equation of state $w_x>-1$ are shown in Tab.~\ref{tab:results-IDEq} and Fig.~\ref{fig:figures-IDEq}.

Notice that the amount of cold dark matter is much lower than within the canonical $\Lambda$CDM scenario. The reason for that is due to the energy flow from the DM to the DE sector, and, consequently, the amount of current dark matter energy density diminishes as $\xi$ increases, see Fig.~\ref{fig:figures-IDEq} and Ref.~\cite{DiValentino:2019ffd,DiValentino:2019jae}. For this very same reason the CMB+R19 data combination prefers a non-zero value of the coupling $\xi$ at a very high significance level. 

For the CMB alone case, see the second column of Tab.~\ref{tab:results-IDEq}, the Hubble constant is shifted towards a higher value, mildly alleviating the Hubble constant tension with R19 at 1.1$\sigma$, mainly via the degeneracy between $\xi$ and $H_0$, as we can notice from Fig.~\ref{fig:figures-IDEq}. When including the BAO data, see the third column of Tab.~\ref{tab:results-IDEq}, most of the parameters are unmodified with respect to the CMB-only case. The tension on $H_0$ with R19 is at the 2.9$\sigma$ level. While the values of $\sigma_8$ and $S_8$ go down with respect to the CMB-only case, they are still far from those corresponding to the $\Lambda$CDM model, even if they are in agreement due to the very large error bars.

When considering the CMB+R19 combination, see the fourth column of Tab.~\ref{tab:results-IDEq}, the Hubble constant value moves in agreement with the R19 measurement within 1$\sigma$, leading to a stronger upper limit on the dark energy equation of state $w_x$ and pushing the value towards the cosmological constant case. In particular, we have $w<-0.904$ at 95\% CL for CMB+R19. As previously stated, we find a mean value of the coupling $\xi$ different from zero at many standard deviations, due to the flux of energy from the DM sector to DE sector. Both $\sigma_8$ and $S_8$ move towards extremely high values with very large error bars, increasing the tension between Planck and the cosmic shear data.
The addition of Pantheon and CC to CMB measurements (fifth column of Tab.\ref{tab:results-IDEq}) leads to a Hubble constant disagreement with R19 at the 3$\sigma$ level. The data set combination case CMB+R19+Pantheon+CC provides constraints which lie between the CMB+Pantheon+CC and CMB+R19 cases, shifting slightly the Hubble constant towards higher values (and therefore reducing the tension with R19 down to the 2.4$\sigma$ level) and setting limits on the coupling $\xi$ different from zero at a high significance level.
\begingroup
\squeezetable
\begin{center}
\begin{table*}
\scalebox{1}{
\begin{tabular}{ccccccccccccccccc}
\hline\hline
Parameters & CMB & CB & CR19 & CPCC & CBPCC & CR19PCC \\ \hline

$\Omega_ch^2$&
$<0.115$&
$0.076_{-0.058}^{+0.046}$&
$<0.055$ & 
$    0.075_{-        0.059}^{+        0.046}$ &
$    0.077_{-        0.059}^{+        0.044}$ & 
$    0.057_{-        0.051}^{+        0.040}$
\\

$\Omega_bh^2$&
$0.02236_{-0.00029}^{+0.00029}$&
$0.02239_{-0.00027}^{+0.00029}$&
$0.02238_{-0.00030}^{+0.00030}$ & 
$    0.02236_{-        0.00028}^{+        0.00029}$ &
$    0.02239_{-        0.00027}^{+        0.00028}$ &
$    0.02241_{-        0.00029}^{+        0.00029}$
\\

$100\theta_{MC}$&
$1.0448_{-0.0043}^{+0.0050}$&
$1.0438_{-0.0031}^{+0.0043}$&
$1.0476_{-0.0029}^{+0.0027}$ & 
$    1.0438_{-        0.0032}^{+        0.0044}$ &
$    1.0437_{-        0.0031}^{+        0.0044}$ &
$    1.0450_{-        0.0032}^{+        0.0042}$
\\

$\tau$&
$0.054_{-0.015}^{+0.016}$&
$0.055_{-0.015}^{+0.017}$&
$0.054_{-0.015}^{+0.015}$ & 
$    0.054_{-        0.015}^{+        0.016}$  &
$    0.055_{-        0.015}^{+        0.016}$ & 
$    0.055_{-        0.015}^{+        0.016}$
\\

$n_s$&
$0.9650_{-0.0084}^{+0.0084}$&
$0.9658_{-0.0084}^{+0.0082}$&
$0.9659_{-0.0082}^{+0.0080}$ & 
$    0.9648_{-        0.0085}^{+        0.0082}$&
$    0.9660_{-        0.0079}^{+        0.0079}$ &
$    0.9661_{-        0.0084}^{+        0.0085}$
\\

${\rm{ln}}(10^{10}A_s)$&
$3.045_{-0.031}^{+0.031}$&
$3.045_{-0.032}^{+0.034}$&
$3.044_{-0.031}^{+0.031}$ & 
$    3.045_{-        0.031}^{+        0.032}$ &
$    3.045_{-        0.031}^{+        0.033}$ & 
$    3.045_{-        0.031}^{+        0.033}$
\\

$w_x$&
$<-0.77$&
$<-0.80$&
$<-0.904$ & 
$<-0.79$ &
$<-0.79$& 
$<-0.834$
\\

$\xi$&
$<0.27$&
$<0.24$&
$0.231_{-0.071}^{+0.060}$ & 
$<0.25$ &
$ <0.24$ & 
 $    0.16_{-    0.10}^{+      0.11}$
\\

$\Omega_{m0}$&
$0.18_{-0.15}^{+0.16}$&
$0.21_{-0.13}^{+0.11}$&
$0.089_{-0.054}^{+0.067}$ & 
$    0.21_{-        0.13}^{+        0.11}$ &
$    0.22_{-        0.13}^{+        0.10}$ & 
 $    0.16_{-       0.10}^{+        0.09}$
\\

$\sigma_8$&
$1.6_{-1.2}^{+2.1}$&
$1.3_{-0.7}^{+1.1}$&
$2.6_{-1.4}^{+1.8}$ & 
$    1.3_{-        0.7}^{+        1.2}$ &
$    1.2_{-        0.6}^{+        1.1}$  & 
$    1.6_{-        0.9}^{+        1.5}$
\\

$H_0{\rm[km/s/Mpc]}$&
$69.3_{-6.5}^{+6.2}$&
$68.4_{-2.5}^{+2.7}$&
$73.3_{-2.3}^{+2.3}$ & 
$   68.2_{-        1.9}^{+       1.9}$ &
$   68.3_{-       1.5}^{+        1.6}$ &
$   70.2_{-       1.6}^{+        1.7}$
\\

$S_8$&
$1.07_{-0.32}^{+0.52}$&
$0.97_{-0.21}^{+0.35}$&
$1.32_{-0.34}^{+0.39}$ & 
$    0.98_{-        0.22}^{+        0.36}$ &
$    0.97_{-        0.20}^{+        0.35}$  &
$    1.06_{-        0.26}^{+        0.43}$
\\

\hline\hline
\end{tabular}
}
\caption{$95\%$~CL constraints within the IDE model in the quintessence-like dark energy scenario ($w_x> -1$) from several dataset combinations, with CB, CR19, CPCC, CBPCC and CR19PCC referring to the combinations of CMB+BAO, CMB+R19, CMB+Pantheon+CC, CMB+BAO+Pantheon+CC and CMB+R19+Pantheon+CC, respectively.}
\label{tab:results-IDEq}
\end{table*}
\end{center}
\endgroup

\subsubsection{IDE + $M_\nu$ - 9 parameters}
\begin{figure*}
\includegraphics[width=0.6\textwidth]{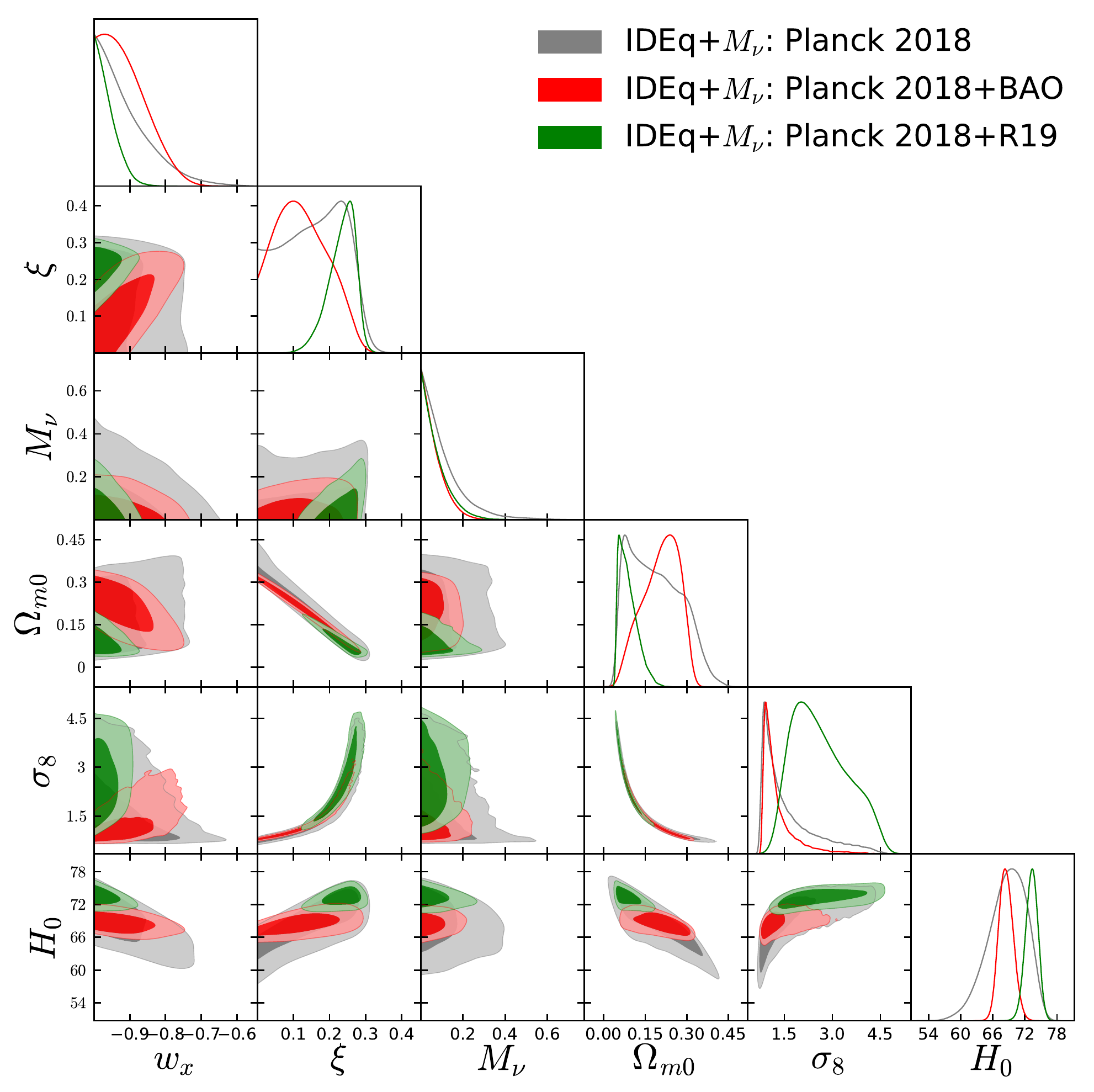}
\caption{68\% and 95\%~CL allowed contours and one-dimensional posterior probability distributions within the IDE$+M_{\nu}$ model in the quintessence-like dark energy scenario ($w_x> -1$) from several data set combinations. Only a sub-set of cosmological parameters are shown.}
\label{fig:figures-IDEqmnu}
\end{figure*}

The results for an IDE + $M_\nu$ scenario, based on nine parameters, with a dark energy equation of state $w_x>-1$ are shown in Tab.~\ref{tab:results-IDEqmnu} and Fig.~\ref{fig:figures-IDEqmnu}.
The addition of a varying neutrino mass is not changing the directions of correlation between the cosmological parameters we have without $M_\nu$, as we can notice from Fig.~\ref{fig:figures-IDEqmnu}. The well-known correlation between $M_\nu$ and the dark energy equation of state $w_x$ is present also in this model, while the negative degeneracy present between $H_0$ and $M_\nu$ is completely absent also in this scenario with $w_x>-1$. 
With respect to the phantom regime, a very mild correlation between $M_\nu$ and $\xi$ appears, especially in the case of the CMB+R19 data set combination, as we can see in Fig.~\ref{fig:figures-IDEqmnu}. For this reason, the $M_\nu$ and $\Omega_\nu h^2$ upper limits improve in CMB+R19 case, contrary to what happens when $w_x<-1$ (see Tab.~\ref{tab:results-IDEpmnu}). Similarly to the phantom model case, the $M_\nu$ upper limits are relaxed, and the most stringent limit we find on the total neutrino mass is for the CMB+Pantheon+CC and CMB+R19+Pantheon+CC data combinations, for which $M_\nu<0.152$~eV at 95\% CL.
\begingroup
\squeezetable
\begin{center}
\begin{table*}
\scalebox{1}{
\begin{tabular}{ccccccccccccccccc}
\hline\hline
Parameters & CMB & CB & CR19 & CPCC & CBPCC & CR19PCC \\ \hline

$\Omega_ch^2$&
$<0.116$&
$0.073_{-0.058}^{+0.048}$&
$<0.056$ & 
$    0.074_{-       0.057}^{+       0.046}$ &
$    0.076_{-        0.058}^{+        0.045}$ & 
$    0.059_{-        0.051}^{+        0.041}$

\\
$\Omega_bh^2$&
$0.02233_{-0.00032}^{+0.00031}$&
$0.02238_{-0.00028}^{+0.00028}$&
$0.02238_{-0.00029}^{+0.00030}$ &
$    0.02234_{-        0.00031}^{+        0.00030}$  &
$    0.02239_{-        0.00028}^{+        0.00028}$ &
$    0.02242_{-        0.00027}^{+        0.00029}$
\\

$100\theta_{MC}$&
$1.0448_{-0.0043}^{+0.0049}$&
$1.0439_{-0.0033}^{+0.0044}$&
$1.0476_{-0.0031}^{+0.0027}$ &
$    1.0438_{-        0.0032}^{+        0.0043}$ &
$    1.0437_{-        0.0031}^{+        0.0043}$ &
$    1.0449_{-        0.0032}^{+        0.0041}$
\\

$\tau$&
$0.054_{-0.015}^{+0.016}$&
$0.055_{-0.015}^{+0.017}$&
$0.054_{-0.015}^{+0.016}$ & 
$    0.054_{-       0.016}^{+        0.016}$ &
$    0.055_{-        0.015}^{+        0.016}$ & 
$    0.055_{-        0.015}^{+        0.016}$
\\

$n_s$&
$0.9644_{-0.0090}^{+0.0089}$&
$0.9660_{-0.0080}^{+0.0079}$&
$0.9657_{-0.0085}^{+0.0084}$ & 
$    0.9646_{-        0.0087}^{+        0.0085}$ &
$    0.9660_{-        0.0080}^{+        0.0081}$ & 
$    0.9665_{-        0.0083}^{+        0.0084}$ 
\\

${\rm{ln}}(10^{10}A_s)$&
$3.045_{-0.031}^{+0.032}$&
$3.045_{-0.031}^{+0.033}$&
$3.044_{-0.032}^{+0.033}$ & 
$    3.045_{-        0.031}^{+        0.032}$ &
$    3.045_{-        0.031}^{+        0.033}$ & 
$    3.045_{-        0.031}^{+        0.033}$
\\

$w_x$&
$<-0.76$&
$<-0.79$&
$<-0.905$ & 
$<-0.80$ &
$<-0.79$  &
$<-0.835$ 
\\

$\xi$&
$<0.28$&
$<0.25$&
$0.234_{-0.081}^{+0.067}$ & 
$ <0.26$ &
$ <0.24$ & 
$    0.158_{-        0.10}^{+        0.11}$
\\

$\Omega_{m0}$&
$0.19_{-0.13-0.15}^{+0.17}$&
$0.21_{-0.13}^{+0.11}$&
$0.089_{-0.054}^{+0.069}$ & 
$    0.21_{-        0.13}^{+        0.11}$ &
$    0.21_{-        0.13}^{+        0.10}$ &
 $    0.17_{-        0.10}^{+        0.09}$
\\

$\sigma_8$&
$1.6_{-1.1}^{+2.0}$&
$1.3_{-0.7}^{+1.2}$&
$2.6_{-1.5}^{+1.8}$ & 
$    1.3_{-        0.7}^{+        1.1}$ &
$    1.2_{-        0.6}^{+        1.1}$ & 
$    1.5_{-        0.8}^{+        1.4}$
\\

$H_0{\rm[km/s/Mpc]}$&
$68.8_{-7.1}^{+6.6}$&
$68.5_{-2.5}^{+2.8}$&
$73.3_{-2.4}^{+2.2}$ & 
$   68.1_{-        2.1}^{+      2.1}$ &
$   68.3_{-       1.5}^{+        1.6}$ &
$   70.2_{-       1.6}^{+        1.7}$
\\

$M_{\nu}$ &
$< 0.318 $&
$< 0.173 $ &
$< 0.203 $ & 
$<0.263$ &
$<0.152$ &
$<0.152$
\\

$\Omega_\nu h^2$&
$<0.0034$&
$<0.0019$&
$<0.0022$ & 
$<0.0028$ &
$<0.0016$  &
$<0.0016$
\\

$S_8$&
$1.08_{-0.31}^{+0.51}$&
$0.99_{-0.22}^{+0.36}$&
$1.33_{-0.35}^{+0.39}$ & 
$    0.98_{-       0.21}^{+        0.33}$ & 
 $    0.97_{-        0.20}^{+        0.34}$ &
 $    1.05_{-        0.25}^{+        0.40}$
\\

\hline\hline
\end{tabular}
}
\caption{$95\%$~CL constraints within the IDE$+M_{\nu}$ model in the quintessence-like dark energy scenario ($w_x> -1$) from several dataset combinations, with CB, CR19, CPCC, CBPCC and CR19PCC referring to the combinations of CMB+BAO, CMB+R19, CMB+Pantheon+CC, CMB+BAO+Pantheon+CC and CMB+R19+Pantheon+CC, respectively.}
\label{tab:results-IDEqmnu}
\end{table*}
\end{center}
\endgroup

\subsubsection{IDE + $N_{\rm eff}$ - 9 parameters}

\begin{figure*}
\includegraphics[width=0.6\textwidth]{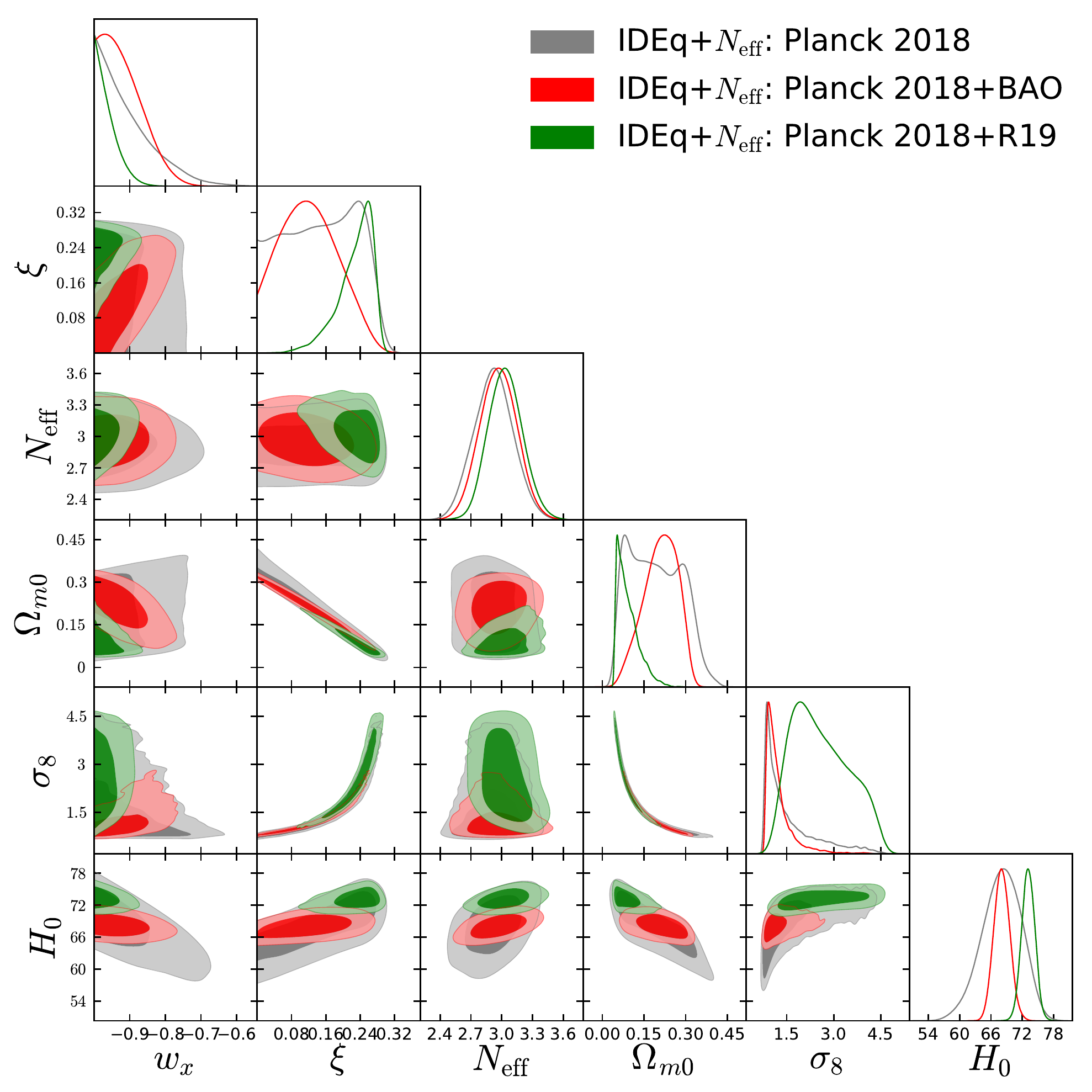}
\caption{$95\%$~CL constraints within the IDE$+N_{\rm eff}$  model in the quintessence-like dark energy scenario ($w_x> -1$) from several dataset combinations, with CB, CR19, CPCC, CBPCC and CR19PCC referring to the combinations of CMB+BAO, CMB+R19, CMB+Pantheon+CC, CMB+BAO+Pantheon+CC and CMB+R19+Pantheon+CC, respectively.}
\label{fig:figures-IDEqneff}
\end{figure*}

Table~\ref{tab:results-IDEqneff} and Fig.~\ref{fig:figures-IDEqneff} depict the results for an IDE + $N_{\rm eff}$ scenario, based on nine parameters. Concerning the correlation between $N_{\rm eff}$ and $n_s$, we notice a shift of 1$\sigma$ of the spectral index towards lower values, followed by a shift of $H_0$ in the same direction.

The effective number of relativistic degrees of freedom $N_{\rm eff}$ is perfectly consistent with its standard value $N_{\rm eff}=3.046$~\cite{Mangano:2005cc,deSalas:2016ztq} for all the data set combinations considered here, with mean values similar to those obtained in the $w_x<-1$ scenario.
A neutrino effective number $N_{\rm eff}$ varying freely is not affecting the directions of correlation between the cosmological parameters, as we can see in Fig.~\ref{fig:figures-IDEqneff}, while the very strong correlation between $N_{\rm eff}$ and the Hubble constant $H_0$ present within the $\Lambda$CDM model is considerably reduced in this interacting scenario. 
The $N_{\rm eff}$ value obtained for the CMB+R19, CMB+Pantheon+CC and CMB+BAO+Pantheon+CC data combinations is very similar and very close to $3$. However, if we consider the CMB+R19+Pantheon+CC data combination (see the last column of Tab.~\ref{tab:results-IDEqneff}), we obtain $N_{\rm eff}=3.23^{+0.30}_{-0.29}$ at 95\% CL.
\begingroup
\squeezetable
\begin{center}
\begin{table*}
\scalebox{1}{
\begin{tabular}{ccccccccccccccccc}
\hline\hline
Parameters &  CMB & CB & CR19 & CPCC & CBPCC & CR19PCC \\ \hline
$\Omega_ch^2$&
$<0.113$&
$0.074_{-0.053}^{+0.046}$&
$<0.065$ & 
$<0.112$ &
$    0.077_{-        0.057}^{+        0.045}$ & 
$    0.064_{-        0.056}^{+        0.047}$
\\

$\Omega_bh^2$&
$0.02223_{-0.00044}^{+0.00044}$&
$0.02231_{-0.00040}^{+0.00041}$&
$0.02237_{-0.00037}^{+0.00040}$ &
$    0.02232_{-        0.00037}^{+        0.00038}$ & 
$    0.02237_{-        0.00036}^{+       0.00037}$ & 
$    0.02254_{-        0.00036}^{+        0.00037}$ 
\\

$100\theta_{MC}$&
$1.0448_{-0.0042}^{+0.0050}$&
$1.0439_{-0.0033}^{+0.0040}$&
$1.0475_{-0.0037}^{+0.0031}$ &
$    1.0439_{-       0.0034}^{+    0.0047}$  &
$    1.0437_{-        0.0032}^{+        0.0043}$ & 
$    1.0445_{-        0.0035}^{+        0.0045}$
\\

$\tau$&
$0.053_{-0.015}^{+0.016}$&
$0.055_{-0.015}^{+0.016}$&
$0.054_{-0.016}^{+0.017}$ & 
$    0.054_{-       0.015}^{+        0.016}$ &
$    0.055_{-       0.015}^{+        0.016}$ & 
$    0.056_{-       0.015}^{+        0.016}$
\\

$n_s$&
$0.960_{-0.018}^{+0.017}$&
$0.963_{-0.015}^{+0.015}$&
$0.965_{-0.014}^{+0.015}$ & 
$    0.963_{-        0.014}^{+        0.014}$ &
$    0.966_{-        0.013}^{+        0.014}$ & 
$    0.973_{-        0.013}^{+        0.013}$
\\

${\rm{ln}}(10^{10}A_s)$&
$3.038_{-0.036}^{+0.037}$&
$3.041_{-0.035}^{+0.036}$&
$3.044_{-0.036}^{+0.037}$ & 
$    3.042_{-       0.034}^{+        0.035}$ &
$    3.044_{-        0.035}^{+        0.036}$ & 
$    3.054_{-        0.034}^{+        0.035}$
\\

$w_x$&
$<-0.76$&
$<-0.81$&
$<-0.902$ &
$<-0.79$ &
$<-0.79$ & 
$<-0.823$
\\

$\xi$&
$<0.27$&
$<0.24$&
$0.229_{-0.090}^{+0.069}$ & 
$    0.13_{-       0.12}^{+    0.13}$ &
$    <0.24$ &
$    0.15_{-        0.11}^{+        0.12}$
\\

$\Omega_{m0}$&
$0.19_{-0.16}^{+0.16}$&
$0.21_{-0.12}^{+0.11}$&
$0.092_{-0.061}^{+0.082}$ &
$    0.21_{-        0.13}^{+        0.11}$ &
$    0.22_{-        0.13}^{+        0.10}$ & 
$    0.17_{-        0.11}^{+        0.10}$
\\

$\sigma_8$&
$1.6_{-1.1}^{+2.0}$&
$1.24_{-0.61}^{+0.99}$&
$2.6_{-1.5}^{+1.8}$ & 
$    1.3_{-        0.7}^{+        1.3}$ &
$    1.2_{-        0.6}^{+        1.0}$ & 
$     1.5_{- 0.8}^{+ 1.5}$ 
\\

$H_0{\rm[km/s/Mpc]}$&
$68.1_{-7.5}^{7.3}$&
$68.1_{-2.9}^{+3.0}$&
$73.2_{-2.5}^{2.4}$ & 
$   68.0_{-        2.5}^{+        2.6}$ &
$   68.2_{-        2.1}^{+        2.1}$ &
$   70.9_{-        1.9}^{+        2.0}$
\\

$N_{\rm eff}$&
$2.92_{-0.37}^{+0.38}$&
$2.97_{-0.35}^{+0.35}$&
$3.04_{-0.31}^{+0.34}$ & 
$    3.00_{-        0.31}^{+        0.32}$ &
$    3.03_{-        0.32}^{+        0.32}$  &
 $    3.23_{-       0.29}^{+        0.30}$
\\

$S_8$&
$1.07_{-0.31}^{+0.51}$&
$0.97_{-0.20}^{+0.31}$&
$1.31_{-0.37}^{+0.40}$ & 
$    0.99_{-        0.22}^{+        0.39}$ &
$    0.96_{-        0.19}^{+        0.32}$ &
$    1.04_{-        0.26}^{+        0.43}$
\\

\hline\hline
\end{tabular}
}
\caption{$95\%$~CL constraints within the IDE$+N_{\rm eff}$ model in the quintessence-like dark energy scenario ($w_x> -1$) from several dataset combinations, with CB, CR19, CPCC, CBPCC and CR19PCC referring to the combinations of CMB+BAO, CMB+R19, CMB+Pantheon+CC, CMB+BAO+Pantheon+CC and CMB+R19+Pantheon+CC, respectively.}
\label{tab:results-IDEqneff}
\end{table*}
\end{center}
\endgroup
\begingroup
\squeezetable
\begin{center}
\begin{table*}
\scalebox{1}{
\begin{tabular}{ccccccccccccccccc}
\hline\hline
Parameters & CMB & CB & CR19 & CPCC & CBPCC & CR19PCC \\ \hline

$\Omega_ch^2$&
$<0.113$&
$0.070_{-0.058}^{+0.049}$&
$<<0.061$ & 
$    0.074_{-       0.055}^{+        0.046}$ & 
$    0.077_{-        0.058}^{+        0.045}$ & 
$    0.064_{-        0.057}^{+        0.046}$
\\

$\Omega_bh^2$&
$0.02222_{-0.00045}^{+0.00046}$&
$0.02231_{-0.00039}^{+0.00040}$&
$0.02238_{-0.00036}^{+0.00039}$ & 
$    0.02232_{-        0.00038}^{+        0.00040}$ & 
$    0.02236_{-        0.00035}^{+        0.00036}$ & 
$    0.02254_{-    0.00036}^{+       0.00036}$
\\
$100\theta_{MC}$&
$1.0448_{-0.0042}^{+0.0049}$&
$1.0442_{-0.0036}^{+0.0045}$&
$1.0476_{-0.0034}^{+0.0030}$ & 
$    1.0439_{-      0.0032}^{+        0.0042}$ & 
$    1.0437_{-        0.0032}^{+        0.0044}$ & 
$    1.0444_{-        0.0035}^{+        0.0046}$ 
\\

$\tau$&
$0.053_{-0.016}^{+0.016}$&
$0.054_{-0.016}^{+0.016}$&
$0.054_{-0.016}^{+0.017}$ & 
$    0.054_{-        0.015}^{+        0.016}$ & 
$    0.055_{-        0.015}^{+        0.016}$ & 
$    0.056_{-        0.015}^{+        0.016}$
\\

$n_s$&
$0.959_{-0.017}^{+0.017}$&
$0.962_{-0.015}^{+0.016}$&
$0.966_{-0.014}^{+0.014}$ & 
$    0.963_{-        0.015}^{+        0.015}$ & 
$    0.965_{-        0.013}^{+        0.014}$ & 
$    0.973_{-        0.013}^{+        0.013}$
\\
${\rm{ln}}(10^{10}A_s)$&
$3.039_{-0.036}^{+0.036}$&
$3.041_{-0.036}^{+0.037}$&
$3.044_{-0.037}^{+0.037}$ & 
$    3.043_{-        0.036}^{+        0.036}$ & 
$    3.044_{-        0.034}^{+        0.035}$ & 
$    3.054_{-        0.034}^{+        0.036}$
\\

$w_x$&
$<-0.75$&
$<-0.80$&
$<-0.90$ & 
$<-0.80$ & 
$<-0.79$ & 
$<-0.818$
\\
$\xi$&
$<0.28$&
$<0.26$&
$0.231_{-0.086}^{+0.068}$ & 
$ <0.25$  & 
$ <0.24$ & 
$    0.15_{-        0.11}^{+        0.12}$
\\
$\Omega_{m0}$&
$0.19_{-0.16}^{+0.16}$&
$0.20_{-0.13}^{+0.12}$&
$0.091_{-0.056}^{+0.075}$ & 
$    0.21_{-        0.12}^{+        0.11}$ & 
$    0.22_{-        0.13}^{+        0.10}$& 
$    0.17_{-        0.11}^{+        0.10}$
\\

$\sigma_8$&
$1.6_{-1.1}^{+1.9}$&
$1.3_{-0.7}^{+1.3}$&
$2.6_{-1.5}^{+1.7}$ & 
$    1.3_{-        0.6}^{+        1.0}$ & 
$    1.2_{-        0.6}^{+        1.0}$ & 
$    1.5_{-        0.9}^{+        1.6}$
\\

$H_0$&
$67.7_{-7.4}^{+7.3}$&
$68.1_{-2.9}^{+3.3}$&
$73.2_{-2.4}^{+2.4}$ &
$   68.0_{-        2.6}^{+        2.7}$ & 
$   68.2_{-        2.1}^{+        2.2}$ & 
$   70.8_{-        1.9}^{+       2.0}$ 
\\
$M_{\nu}$&
$< 0.337 $&
$< 0.170 $&
$< 0.199 $ & 
$<0.243$ & 
$<0.156$ & 
$<0.158$
\\

$N_{\rm eff}$&
$2.92_{-0.37}^{+0.38}$&
$2.96_{-0.35}^{+0.37}$&
$3.05_{-0.32}^{+0.33}$ & 
$    3.01_{-       0.32}^{+        0.33}$ & 
$    3.02_{-        0.30}^{+        0.33}$ & 
$    3.22_{-        0.29}^{+        0.29}$
\\

$\Omega_\nu h^2$&
$<0.0035$&
$<0.0018$&
$<0.0021$ & 
$<0.0026$ & 
$<0.0017$ & 
$<0.0017$
\\
$S_8$&
$1.06_{-0.30}^{+0.49}$&
$1.00_{-0.23}^{+0.39}$&
$1.32_{-0.38}^{+0.40}$ & 
$    0.98_{-        0.20}^{+        0.32}$ & 
$    0.96_{-        0.20}^{+        0.33}$ & 
$    1.04_{-        0.26}^{+        0.43}$
\\

\hline\hline
\end{tabular}
}
\caption{$95\%$~CL constraints within the IDE$+M_{\nu}+N_{\rm eff}$ model in the quintessence-like dark energy scenario ($w_x> -1$) from several dataset combinations, with CB, CR19, CPCC, CBPCC and CR19PCC referring to the combinations of CMB+BAO, CMB+R19, CMB+Pantheon+CC, CMB+BAO+Pantheon+CC and CMB+R19+Pantheon+CC, respectively.}
\label{tab:results-IDEqmnuneff}
\end{table*}
\end{center}
\endgroup
\begin{figure*}
\includegraphics[width=0.6\textwidth]{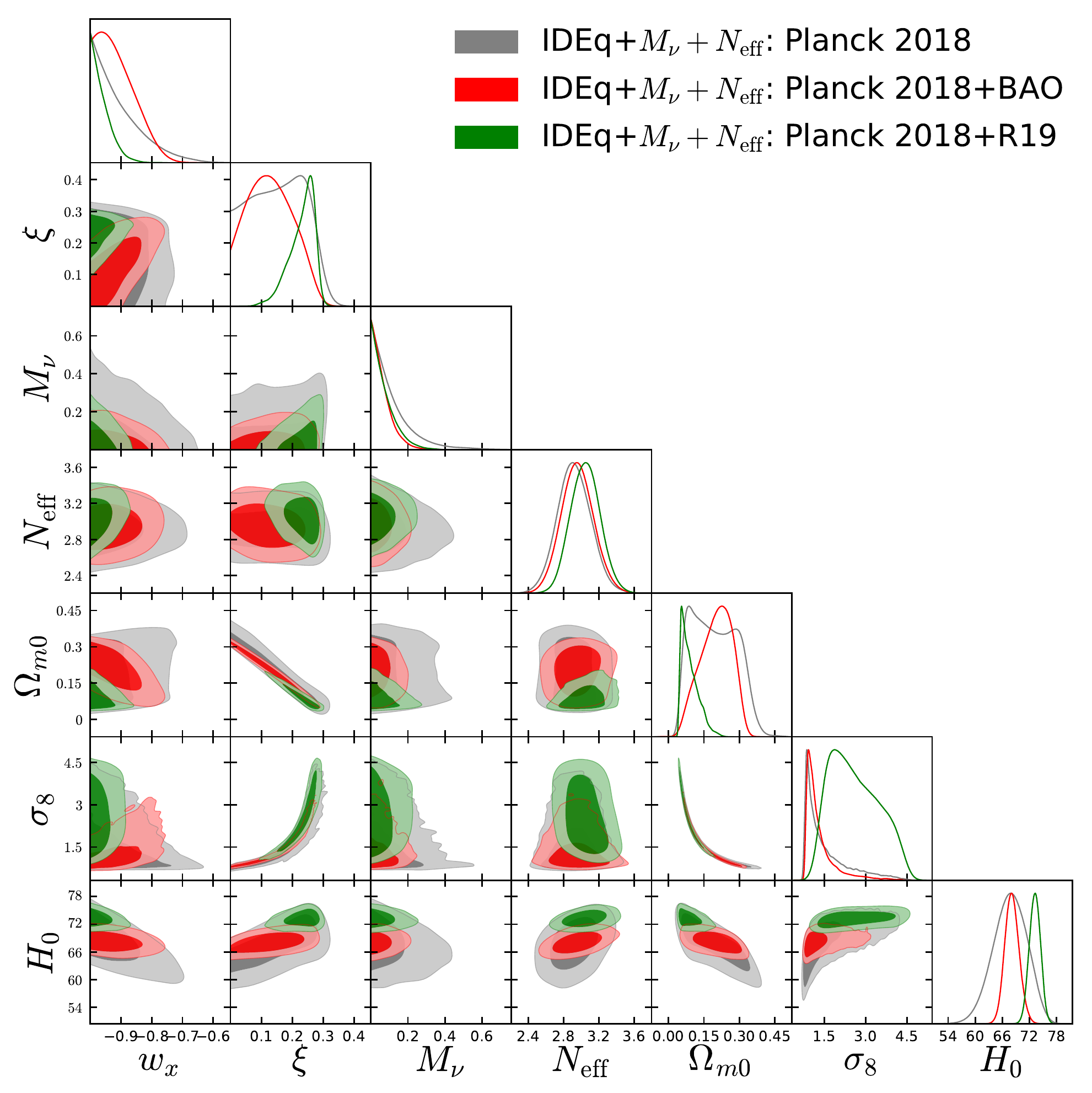}
\caption{$95\%$~CL constraints within the IDE$+M_{\nu}+N_{\rm eff}$ in the quintessence-like dark energy scenario ($w_x> -1$) from several dataset combinations, with CB, CR19, CPCC, CBPCC and CR19PCC referring to the combinations of CMB+BAO, CMB+R19, CMB+Pantheon+CC, CMB+BAO+Pantheon+CC and CMB+R19+Pantheon+CC, respectively.}
\label{fig:figures-IDEqmnuneff}
\end{figure*}
                       
\subsubsection{IDE + $M_\nu$ + $N_{\rm eff}$ - 10 parameters}

The results for this scenario are shown in Tab.~\ref{tab:results-IDEqmnuneff} and Fig.~\ref{fig:figures-IDEqmnuneff}. In this full scenario, where both the total neutrino mass $M_\nu$ and the effective number of relativistic degrees of freedom $N_{\rm eff}$ are varying simultaneously, both the parameters $n_s$ and $H_0$ are shifted to lower values when compared to the previous IDE+$N_{\rm eff}$ model shown in Tab.~\ref{tab:results-IDEqneff}.

The upper limits on the total neutrino mass $M_\nu$ are only mildly relaxed with respect to the IDE+$M_\nu$ model showed in Tab.~\ref{tab:results-IDEpmnu} for the CMB-only case, while are slightly stronger for the CMB+Pantheon+CC data set combination. The most stringent upper limit we have on this parameter is obtained here for the data set combination CMB+BAO+Pantheon+CC (see the sixth column of Tab.~\ref{tab:results-IDEqmnuneff}) for which $M_\nu<0.156$~eV at 95\% CL.

\section{Summary}
\label{sec-5}

In this paper we have explored possible extensions of the Interacting Dark Energy, where the dark energy and the dark matter fluids interact with each other. We have focused on the impact of such interactions on the dark radiation sector, allowing the neutrino mass and the effective number of neutrino species free to vary, both individually and simultaneously. The effect of such a dark coupling on the dark radiation sector is analyzed within two different Dark Energy regimes: a phantom ($w_x<-1$) and a quintessence ($w_x>-1$) scenario. We have exploited the most recent publicly available cosmological observations, which include the Planck 2018 legacy data, Baryon Acoustic Oscillations, the most recent measurement of the Hubble constant using the Cepheids as calibrators, Supernovae Type Ia Pantheon data and measurements of the Hubble parameter from Cosmic Chronometers. We find, in general, that the constraints on the dark radiation sector physics are quite close to those found within the minimal $\Lambda$CDM cosmology. The derived bounds are almost independent of the dark energy regime (phantom versus quintessence), contrary to the case without a dark coupling~\cite{Yang:2017rcn, Vagnozzi:2018jhn}. We find a total neutrino mass $M_\nu<0.15$~eV and a number of effective relativistic degrees of freedom of $N_{\rm eff}=3.03^{+0.33}_{-0.33}$, both at 95\%~CL, which are indeed not that far from those obtained within the $\Lambda$CDM cosmology, $M_\nu<0.12$~eV and $N_{\rm eff}=3.00^{+0.36}_{-0.35}$ for the same data combination. 
Current cosmological observations are therefore powerful enough to disentangle the physical effects associated to the different dark sector components, i.e. dark energy, dark matter and dark radiation.

On the other hand, it might be interesting to investigate the mutual interacting scenario in the full dark sector of the universe, namely, between Dark energy-Dark matter-Dark radiation, considering different rates of interaction, and examine how such coupling can affect both the dark and neutrino sector in light of the recent cosmological observations.

\section{Acknowledgments}

WY acknowledges the support from the 
National Natural Science Foundation of China under Grants 
No. 11705079 and No. 11647153. EDV was supported from the European Research Council in the form of a Consolidator Grant 
with number 681431. OM is supported by the Spanish grants FPA2017-85985-P and SEV-2014-0398 of the MINECO, by PROMETEO/2019/083 and by the European Union Horizon 2020 research and innovation program (grant agreements No. 690575 and 67489). SP has been supported by the Mathematical Research Impact-Centric Support Scheme  (MATRICS), File No. MTR/2018/000940, given by the Science and Engineering Research Board (SERB), Govt. of India. 
RCN would like to thank the FAPESP for financial support under the project \# 2018/18036-5.

%--------------------------------------------------------------

%%%%%%%%%%%%%%%%%%%%%%%%%%%%%%%%%%%%%%%%%%%%%%%%%%%%%%%%%%%%%%%%%%%%%%%%%%%%%%%%%%%%%%%%%%%%%%%%%%%%%%%%
%%%%%%%%%%%%%%%%%%%%%%%%%%%%%%%%%%%%%%%%%%%%%%%%%%%%%%%%%%%%%%%%%%%%%%%%%%%%%%%%%%%%%%%%%%%%%%%%%%%%%%%%%
\end{document}